\documentclass[fleqn,1p]{elsarticle}

\usepackage{hyperref}

\usepackage{graphicx}
\journal{Journal}
\usepackage{booktabs}
\usepackage{tikz}
\usepackage{xcolor}
\usepackage[nameinlink,noabbrev]{cleveref}
\usepackage{rotating}
\usepackage{amssymb}
\usepackage{tablefootnote}
\usepackage{svg}
\usepackage{longtable}
 \usepackage{booktabs} 

\hypersetup{colorlinks = true, allcolors = black}
 \usetikzlibrary{trees,positioning,shapes,shadows,arrows}

\begin{document}
\tikzset{
  basic/.style  = {draw, text width=2cm, drop shadow, font=\sffamily, rectangle},
  root/.style   = {basic, rounded corners=4pt, thin, align=center, fill=white, text width=3cm},
  level-2/.style = {basic, rounded corners=6pt, thin,align=center, fill=white, text width=2cm},
  level-3/.style = {basic, thin, align=center, fill=white, text width=2cm}
}
\begin{frontmatter}

\title{Machine Learning for Stuttering Identification: Review, Challenges and Future Directions}



\author[mysecondaryaddress]{Shakeel A.~ Sheikh\corref{corr1}}
\ead{shakeel-ahmad.sheikh@loria.fr}

\author[mysecondaryaddress]{Md Sahidullah}
\ead{md.sahidullah@inria.fr}

\author[mymainaddress]{Fabrice Hirsch}
\ead{fabrice.hirsch@univ-montp3.fr}

\author[mysecondaryaddress]{Slim Ouni}
\ead{slim.ouni@loria.fr}

\address[mysecondaryaddress]{Universit\'{e} de Lorraine, CNRS, Inria, LORIA, F-54000, Nancy, France}

\address[mymainaddress]{Université Paul-Valéry Montpellier, CNRS, Praxiling, Montpellier, France}

\cortext[corr1]{Corresponding author}





\begin{abstract}
Stuttering is a speech disorder during which the flow of speech is interrupted by involuntary pauses and repetition of sounds. Stuttering identification is an interesting interdisciplinary domain research problem involving pathology, psychology, acoustics, and signal processing, making it hard and complicated to detect. Recent developments in the machine and deep learning have dramatically revolutionized the speech domain, however minimal attention has been given to stuttering identification. This work fills the gap by trying to bring researchers together from interdisciplinary fields. In this paper, we comprehensively review acoustic features, and statistical and deep learning-based stuttering/disfluency classification methods. We also present several challenges and possible future directions. 
\end{abstract}


\end{frontmatter}


\section{Introduction}

\label{introduction}
Speech disorders or speech impairments are communication disorders in which a person has difficulties in creating and forming the normal speech sounds required to communicate with others~\cite{guitar2013stuttering, duffy2013motor}. These disorders can take the form of dysarthria, apraxia, stuttering, cluttering, lisping, and so on~\cite{duffy2013motor, ratner2018fluency, guitar2013stuttering, ward2008stuttering, kehoe2006speech}. 
\par 

\emph{Dysarthria} is defined as a speech disorder caused by muscle weakness (including face, lips, tongue, and throat) controlled by the nervous system. The \emph{patients with dysarthria} produce slurred or mumbled sounds with aberrant speech patterns, such as flat intonation or very low or fast speech rate, which makes their speech very difficult to comprehend~\cite{duffy2013motor}. \emph{Cluttering} is characterized by a patient's speech being too jerky, too rapid, or both. \emph{Persons with cluttering} usually exclude/collapse most of the syllables, or aberrant rhythms or syllable stresses, and also contain excessive amounts of interjections such as {\itshape so}, {\itshape} hmm, {\itshape like}, {\itshape umm,} etc~\cite{ward2008stuttering}. \emph{Apraxia} is defined as a speech disorder when the neural path between the nervous system and the muscles responsible for speech production is obscured or lost. The \emph{persons with apraxia} knows what they want to speak, but can not speak because the brain is unable to send an exact message to the speech muscles which can articulate the intended sounds, despite the fact speech muscle movements are working fine~\cite{duffy2013motor, kehoe2006speech}. 

\emph{Lisping} speech disorder is defined as the incapability of producing sibilant consonants (\textit{z or s}) correctly. The sibilant sounds are usually substituted by {\itshape th} sounds. For example, the persons with lisping speech disorder would pronounce the word {\itshape lisp} as {\itshape lithp}~\cite{duffy2013motor}. Stuttering speech impairment is different than other speech disorders because it can be cured if early intervention is being made to help the persons who stutter (PWS) develop normal fluency~\cite{guitar2013stuttering}. Of all these speech impairments, stuttering is the most common one\footnote{\url{https://www.healthline.com/health/speech-disorders}}.
\par 
Stuttering~-~also called stammering/disfluency\footnote{In this review, we will use the terms disfluency, stuttering, and stammering interchangeably}~-~ is a neuro-developmental speech disorder that commences when neural connections supporting language, speech, and emotional functions are quickly changing~\cite{duffy2013motor, ward2008stuttering, smith2017stuttering}, and is characterized by core behaviors that usually take the form of involuntary stoppages, repetition, and prolongation of sounds, syllables, words or phrases. Stuttering can be described as an abnormally and persistent duration of stoppages in the normal forward flow of speech \cite{guitar2013stuttering}. These speech abnormalities are generally accompanied by unusual behaviors like head nodding, lip tremors, quick eye blinks, unusual lip shapes, etc \cite{riva2008phenomenology}. Fluency can be defined as the capacity to produce speech without any effort, at a normal rate~\cite{starkweather1987fluency}. A fluent speech requires linguistic knowledge of the spoken language and a mastery of the message content. Concerning physiological aspects, a precise respiratory, laryngeal and supraglottic control movement is necessary to maintain fluency~\cite{adams1974physiologic}. When all these conditions are not met, speech disorder (stuttering) can emerge. They can take the form of silent or filled pauses, repetitions, interjections, revisions (content change or grammatical structure or pronunciation change), and incomplete phrases~\cite{roberts2009disfluencies}. Generally, normal speech is made up of mostly the fluent and some disfluent parts. Notice that normal disfluencies are useful in speech production since they can be considered in time during which the speaker can correct or plan the upcoming discourse. In some cases, like stuttering, disfluencies do not help the speaker to organize his/her discourse. Indeed, contrary to persons without any fluency disorder, PWS know what they want to pronounce but are momentarily unable to produce it~\cite{world1977manual, guitar2013stuttering}. 
\par 
Stuttering can broadly be classified into two types~\cite{guitar2013stuttering}:
\begin{itemize}
    \item \textit{Developmental Stuttering:} This stuttering is the most common one and it usually occurs in the learning phase of the language, \textit{i.e. between two and seven}. Recent researches conclude that developmental stuttering is multifactorial trouble including neurological and genetic aspects \cite{etchell2018systematic, drayna2011genetic}. Indeed, fMRI studies show anomalies in neural activity before the speech, \textit{i.e.} during the planning stages of speech production~\cite{vanhoutte2016will}. Furthermore, an atypical activation in the left inferior frontal gyrus and right auditory regions~\cite{neef2015speech,belyk2015stuttering} has been highlighted. Concerning the genetic aspects, Riaz \emph{et al.}~\cite{riaz2005genomewide} observe an unusual allele on chromosom 12 by PWS. Drayna and Kang~\cite{drayna2011genetic} identify 87 genes that could be involved in stuttering, including one called GNPTAB, which was significantly present in a lot of PWS.
    
    \item \textit{Neurogenic Stuttering:} This stuttering can occur after head trauma, brain stroke, or any kind of brain injury. This results in disfluent speech because of the incoordination of the different regions of the brain which are involved in speaking \cite{stuttering}. Even if neurogenic stuttering is rare, it can be observed in children and adults regardless of their age.
\end{itemize}
Globally, stuttering concerns 1\% of the world’s population and its incidence rate is between 5\% and 17\% \cite{yairi2013epidemiology}. The difference between the prevalence and incidence rates can be explained by the fact that developmental stuttering disappears in 80\% of the cases before adulthood either without any intervention or thanks to speech therapy. Thus, about 70 million people suffer from this trouble which affects four times males than females~\cite{yairi2013epidemiology}. As considered by the non-stuttering persons, the disfluency affects the flow of speech only, however for PWS, it is greater than that. 
Several studies show that PWS are ignored, teased, and/or bullied by normo-fluent people ~\cite{iverach2016prevalence}. The PWS is usually rated less popular than their non-stuttering peers and less likely to be considered leaders~\cite{iverach2016prevalence}. According to the national stuttering association~\cite{national2009experience}, 40\% of the PWS have been repudiated school opportunities, promotions or job offers and it affects relationships. The data should be assessed in close conjunction with the fact that 85\% of businessmen consider stuttering as a negative element during a job interview and prefer offering work that does not require a customer contact~\cite{klein2004impact}. All these elements explain that PWS develop social anxieties and negative feelings (fear, shame, etc.) when they have to speak in public~\cite{blood2016long}.

\par 
Stuttering appears to be complex and mysterious. Several factors that lead to stuttering include stress, delayed childhood development, and speech motor control abnormalities, as there is a strong correlation between stress, anxiety, and stuttering~\cite{guitar2013stuttering}. Indeed, disfluencies are more frequent in stress or anxiety conditions, in dual tasks including speech and another cognitive charge, and when they speak fast. Conversely, PWS produce fewer disfluencies when they sing in unison or speak with an altered auditory feedback~\cite{antipova2008effects}. In a recent study, Smith, and Weber~\cite{smith2017stuttering}, postulated the multifactorial dynamic pathways theory, where they asserted that stuttering occurs because of the failure of the central nervous system in generating the necessary patterns of motor commands for fluent speech. Thus, stuttering shows impairment in sensorimotor processes that are responsible for speech production, and its orientation throughout the life of PWS is strongly affected by linguistic and emotional aspects.

In conventional stuttering assessment, the speech-language pathologists (SLP) or speech therapists (ST)
manually analyze either the PWS’ speech or their recordings~\cite{noth2000automatic, guitar2013stuttering}. The stuttering severity is usually measured
by taking the ratio of disfluent words/duration to the total words/duration~\cite{guitar2013stuttering}. The most conventional
speech therapy sessions involve helping the PWS observe and monitor their speech patterns to
rectify them~\cite{guitar2013stuttering}. The speech therapeutic success rate recoveries have been reported to be 60-80\% when dealt
with in the early stage~\cite{saltuklaroglu2005effective}. This convention of detecting stuttering severity and its improvement due to therapeutic
sessions is very demanding and time-consuming and is also biased and prejudiced towards the subjective belief of SLPs. Due to the nature of stuttering, its therapeutic sessions are very intense course, that usually, extend to several months (several years in some cases), which necessitates PWS to see the SLP regularly~\cite{roberts2011using}. Usually, the speech therapy sessions are private and are very expensive, thus making it unaffordable to some PWS. Thus, it is important to develop interactive automatic SD systems.
\par
The automatic speech recognition systems (ASR) are working well for fluent speech, but they fail to recognize the stuttered speech. So, it would not be feasible for a PWS to easily access virtual assistant tools like Alexa, Apple Siri, etc. \cite{usatodaytech2020}. The SD may help in adapting the virtual assistant tools for disfluent persons. 
\par 
Therefore, automatic stuttering identification systems (ASIS) are the need for an hour which provides an objective and consistent measurement of the stuttered speech. Consequently, with the recent developments in natural language processing, machine learning, and deep learning, it became a reality to develop smart and interactive SD and therapy tools \cite{kourkounakis2020detecting,sheikh:hal-03227223, stutternetmtl, bayerw2v2}. In spite of the fact, that there are numerous applications of ASIS, very little attention has been given to this field.

\begin{figure}
     \includegraphics[scale=0.2]{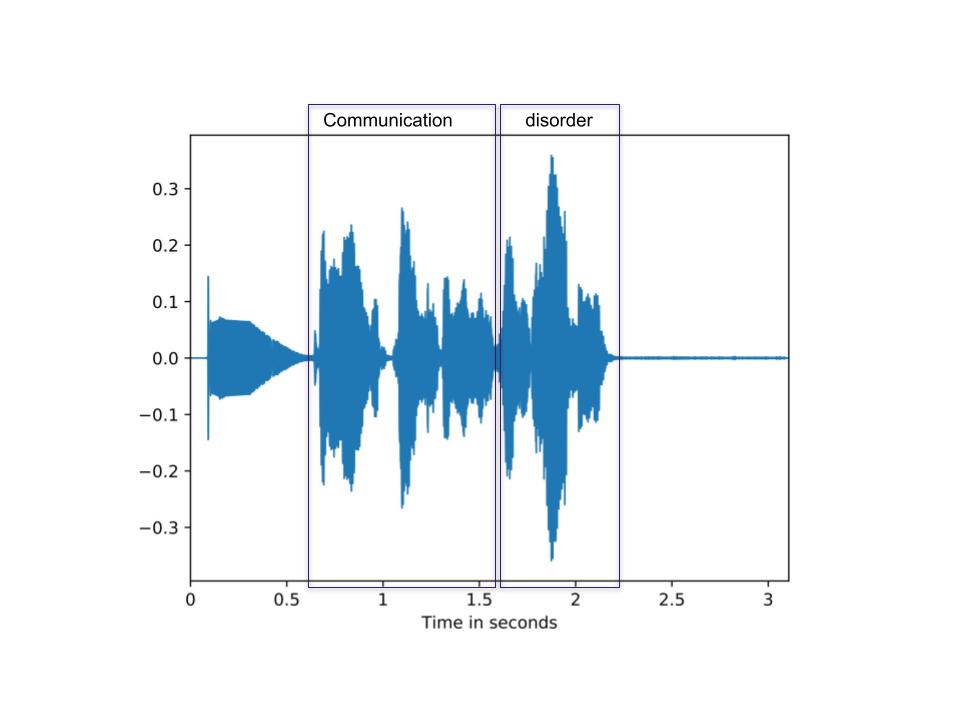} \hfill
     \includegraphics[scale=0.2]{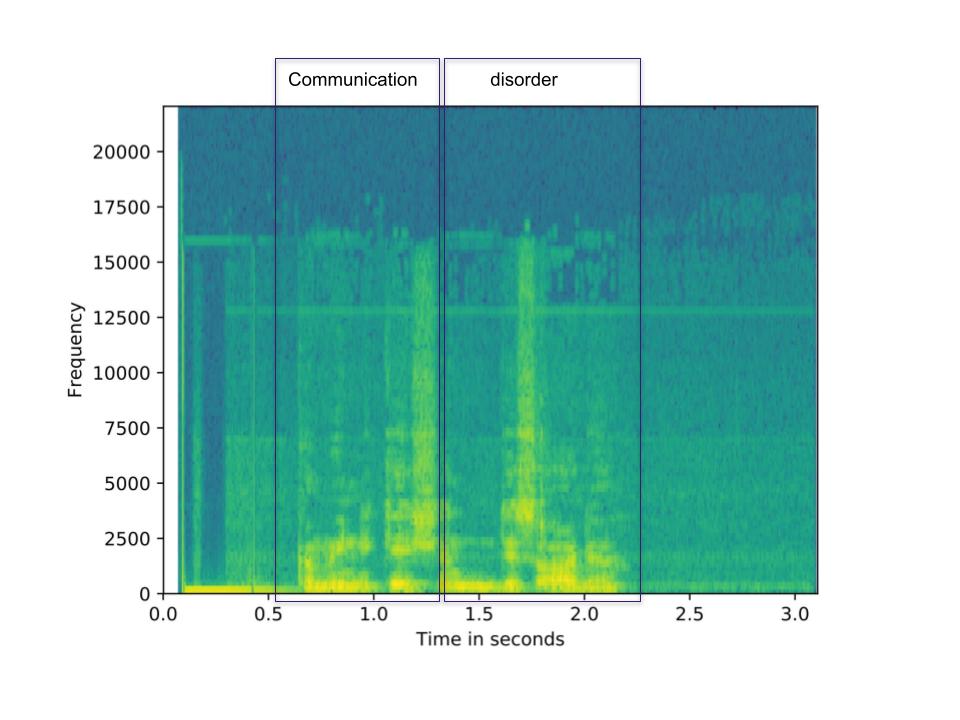} \\ 
     \includegraphics[scale=0.2]{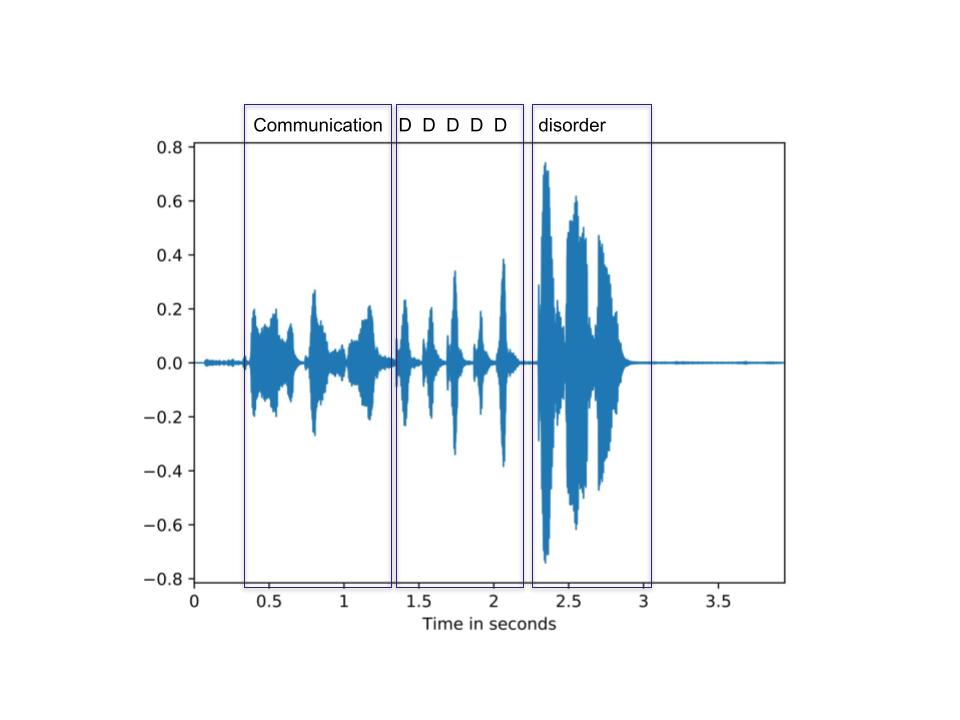} \hfill
     \includegraphics[scale=0.2]{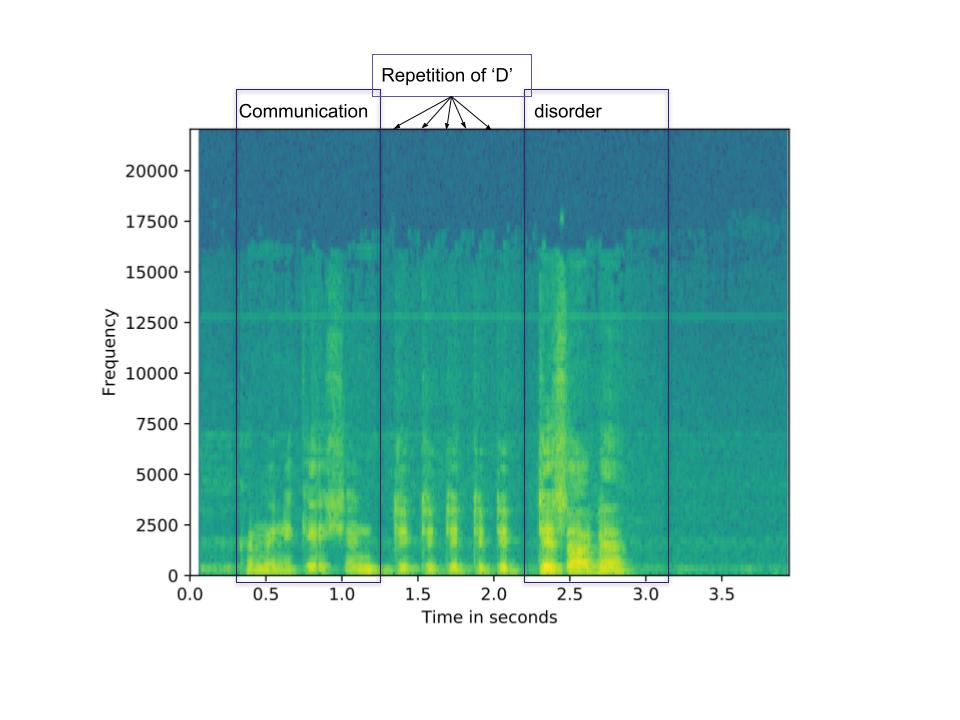} 
     
    \caption{Speech waveforms and spectrograms of a speaker (male) saying \emph{Communication disorder}. The left is waveforms (amplitude v/s time); the right  is a time–frequency plot using a wavelet decomposition of these data. Top row is fluent speech; bottom row is stuttering (repetitions), occur at the  “D” in “disorder.” Five repetitions can be clearly identiﬁed by arrows in the spectrogram. (bottom right)} . 
 \label{fig:spec}
\end{figure}

 \begin{table*}
    \centering
    \scalebox{0.7}{\begin{tabular}{c c c}
        \toprule
        Stutter Type & Definition & Example  \\
         \midrule 
        \midrule
        Blocks & Involuntary pause before a word & I w~\textbf{blockage/pause}~ant to speak \\
        Prolongations&Prolonged Sounds&\textbf{Sssssss}am is kind \\
        Interjection&Insertion of sounds& uh, uhm \\
        Sound Repetition&Phoneme repetition&He \textbf{w-w-w}-wants to write \\
        Part-Word Repetition&Repetition of a Syllable&\textbf{Go-go-go} back \\
        Word Repetition &Repetition of a Word&\textbf{Well, well}, I didn't get you \\
        Phrase Repetition&Repetition of several successive words&\textbf{I have, I have} an iphone\\
        Repetition$-$Prolongation&Repetition and Prolongation&\textbf{Ggg}o \textbf{b-b-b}back \\
        &disfluencies occurring at the same time&\\
        Multiple&Multiple disfluencies in a word or phrase&\textbf{Tt}-\textbf{Tt}-\textbf{Tt}ariq \textbf{blockage/pause} is \textbf{kkkk}ind \\
        False Start& Revision of a phrase or a word &\textbf{I had-} I lost my watch \\
        \bottomrule
    \end{tabular}}
    \caption{Various stuttering types}
    \label{tab:stutter_types}
\end{table*}
\begin{figure}
\centering
\begin{tikzpicture}[
  level 1/.style={sibling distance=8em, level distance=5em},
  edge from parent/.style={->,solid,black,thick,sloped,draw}, 
  edge from parent path={(\tikzparentnode.south) -- (\tikzchildnode.north)},
  >=latex, node distance=0.6cm, edge from parent fork down]

\node[root, text=blue] {\textbf{Overview of ASIS}}
  child {node[level-2,text=blue] (c1) {\textbf{Datasets}}}
  child {node[level-2,text=blue] (c2) {\textbf{Processing}}}
  child {node[level-2,text=blue] (c3) {\textbf{Features}}}
  child {node[level-2,text=blue] (c4) {\textbf{Modalities}}}
  child {node[level-2,text=blue] (c5) {\textbf{Classifiers}}};


\begin{scope}[every node/.style={level-3}]
\node [below of = c1,text=blue, xshift=10pt] (c11) {UCLASS};
\node [below of = c11,text=blue,] (c12) {LibriStutter};
\node [below of = c12,text=blue,] (c13) {Torgo};
\node [below of = c13,text=blue,] (c14) {FluencyBank};
\node [below of = c14,text=blue,] (c15) {SEP-28k};
\node [below of = c15,text=blue,] (c16) {KSoF};

\node [below of = c2,text=blue, xshift=10pt] (c21) {Framing};
\node [below of = c21,text=blue] (c22) {Windowing};
\node [below of = c22,text=blue] (c23) {DCT};
\node [below of = c23,text=blue] (c24) {Filters};

\node [below of = c3,text=blue, xshift=10pt] (c31) {MFCCs};
\node [below of = c31,text=blue] (c32) {Spectrogram};
\node [below of = c32,text=blue] (c33) {LPCC};
\node [below of = c33,text=blue] (c34) {LFPC};
\node [below of = c34,text=blue] (c35) {LH-MFCC};
\node [below of = c35,text=blue] (c36) {PLP};
\node [below of = c36,text=blue] (c37) {Formants};
\node [below of = c37,text=blue] (c38) {Jitter};
\node [below of = c38,text=blue] (c39) {Shimmer};
\node [below of = c39,text=blue] (c310) {Pitch};
\node [below of = c310,text=blue] (c311) {Energy};
\node [below of = c311,text=blue] (c312) {i-vectors};
\node [below of = c312,text=blue] (c313) {Duration};
\node [below of = c313,text=blue] (c314) {ECAPA};
\node [below of = c314,text=blue] (c315) {SOM};
\node [below of = c315,text=blue] (c316) {Wav2Vec2.0};
\node [below of = c316,text=blue] (c317) {ZCR};
\node [below of = c317,text=blue] (c318) {PP};

\node [below of = c4,text=blue, xshift=10pt] (c41) { Audio};
\node [below of = c41,text=blue] (c42) {H.Pulses};
\node [below of = c42,text=blue] (c43) {Textual};
\node [below of = c43,text=blue] (c44) {Visual};

\node [below of = c5, ,text=blue, xshift=10pt] (c51) {SM};
\node [below of = c51,text=blue] (c52) {DLM};

\node [below of = c51,,text=blue,xshift = 10,  yshift=-85pt] (c512) {MLP};
\node [below of = c512,text=blue] (c513) {SVM};
\node [below of = c513,text=blue] (c514) {HMM};
\node [below of = c514,text=blue] (c515) {GMM};
\node [below of = c515,text=blue] (c516) {Correlation};
\node [below of = c516,text=blue] (c517) {ANN};
\node [below of = c517,text=blue] (c518) {$k$-NN};
\node [below of = c518,text=blue] (c519) {LDA};
\node [below of = c519,text=blue] (c5110) {NBC};
\node [below of = c5110,text=blue] (c5111) {SOM};
\node [below of = c5111,text=blue] (c5112) {Fuzzy Logic};
\node [below of = c5112,text=blue] (c5113) {Rough Sets};

\node [below of = c52,,text=blue, xshift = -60, yshift=-65] (c521) {DBF};
\node [below of = c521,text=blue] (c522) {CNN};
\node [below of = c522,text=blue] (c523) {ResNet};
\node [below of = c523,text=blue] (c524) {LSTM};
\node [below of = c524,text=blue] (c525) {Attention};
\node [below of = c525,text=blue] (c526) {FluentNet};
\node [below of = c526,text=blue] (c527) {\emph{StutterNet}};
\node [below of = c527,text=blue] (c528) {\emph{MBStutterNet}};

\end{scope}

\foreach \value in {1,2,3,4,5, 6}
  \draw[->,black] (c1.195) |- (c1\value.west);
\foreach \value in {1,2,3,4}
  \draw[->,black] (c2.195) |- (c2\value.west);
\foreach \value in {1,2,3,4,5,6,7,8,9,10,11,12,13,14,15,16,17, 18}
  \draw[->,black] (c3.195) |- (c3\value.west);
\foreach \value in {1,2,3,4}
  \draw[->,black] (c4.195) |- (c4\value.west);
\foreach \value in {1,2}
  \draw[->,black] (c5.195) |- (c5\value.west);
\foreach \value in {2,3,4,5,6,7,8,9,10,11,12,13}
     \draw[-latex,black] (c51.720) to [bend left=18] (c51\value.east);
\foreach \value in {1,2,3,4,5,6,7, 8}
     \draw[-latex,black] (c52.190)  to [bend right=65] (c52\value.west);

\end{tikzpicture}
    \caption{Overview of automatic stuttering identification systems}
    \small
    DLM: Deep learning models \hspace{1cm} SM: Statistical models \\
               \textit{MLP: Multi layer perceptron, \hfill SVM : Support vector machines \\
               CNN: Convoluitonal neural network, \hfill HMM: Hidden Markow models  \\
               RNN: Recurrent neural network, \hfill GMM: Gaussian mixture models  \\
               LSTM: Long short-term memory, \hfill LDA: Linear discriminant analysis \\
               DBF: Deep belief neural network, \hfill NBC: Naive Bayes classifier \\
               Multi-branch (MB) \emph{StutterNet}, \hfill  SOM: Self organizing maps \\
               \textit{Features} \\
               MFCCs: Mel-frequency cepstral coefficients \\
               LPCC: Linear prediction cepstral coefficients \\
               LFPC: Log frequency power coefficients \\
               DCT: Discrete cosine transforms \\
               PP: Phoneme probabilities}
               
    \label{fig:class}
\end{figure}
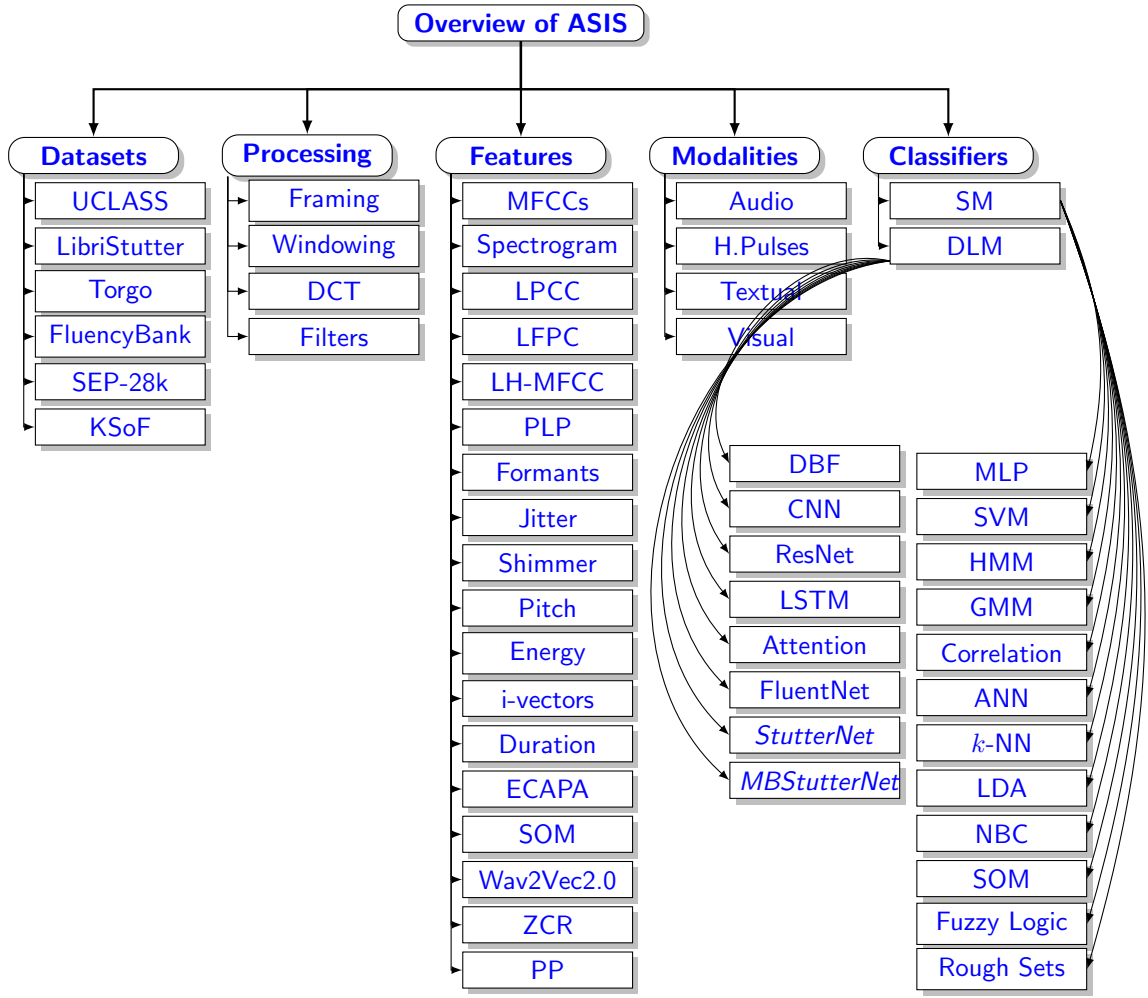
\par 
We define an ASIS as a compilation of techniques and methodologies that takes audio speech signals as an input, pre-processes, and categorizes them to identify the stuttering embedded in them. When we take a broad view of ASIS, we can express it into several domains as shown in Fig.~\ref{fig:class}. It would be extremely useful to understand stuttering better in order to enhance the stuttering classification process. The stuttering problem is still an open problem and it has been approached through several techniques, most of them fall in the supervised learning category~\cite{sheikh:hal-03227223, kourkounakis2020fluentnet, sep28k, stutternetmtl, bayerw2v2, sheikhw2v2}. An ASIS system that consists of a classifier and a supervised learning loss function is trained on the data to recognize and identify stuttering types embedded in the audio speech signal. These supervised learning systems require stuttering embedded labeled data. To feed the data to the model, requires some preprocessing to extract useful features like Mel-frequency cepstral coefficients (MFCCs) which reduces the original data into its important characteristics that are essential for classification purposes. In speech, these can be grouped into spectral, voice, and prosodic features. The spectral ones are mostly used in the literature. In addition to these, features from other modalities such as linguistic (textual) can also be incorporated to improve classification performance. 
\par 
Different researchers have used different acoustic features and classifiers for SD. As such, there is no systematic study that mentions the detailed analysis and challenges of various SD methods. In this paper, we give an up-to-date comprehensive survey by unifying all the stuttering methods. We also give a basic summary of various acoustic characteristics of stuttering, that can be exploited in SD. This paper will be a valuable resource for researchers, engineers, and practitioners. This paper discusses some stuttering datasets that can be referenced by various researchers. Despite the recent advancements by deep learning (DL) in speech recognition~\cite{nassif2019speech}, emotion detection~\cite{akccay2020speech}, speech synthesis\cite{ning2019review}, etc, it has not been much used in the stuttering domain. Due to the data scarcity in the stuttering domain, the DL has not been explored to a great extent in SD. We recently proposed \emph{StutterNet} trained on a small dataset, that shows promising results in SD. In this paper, we also present several challenges faced by various SD methods and their possible solutions to show how we can efficiently use DL to boost SD from the speech utterance. 
\par 
There are several modalities that are considered in SD, which include: speech~\cite{sheikh:hal-03227223}, visual~\cite{yildirim2009automatic}, text~\cite{geetha2000classification}, bio-respiratory signals~\cite{villegas2019novel}, and functional near-infrared spectroscopy~\cite{fmri, chang2014research}. This paper focuses mainly on the speech modality. Most of the stuttering modalities are very expensive to collect besides speech modality because that is cheap and can be collected remotely by a simple mobile application or a web interface. 

\par 
The remainder of the paper is organized as follows. Section \ref{acousticcharecteristics} discusses the various acoustics properties of stuttered speech. This section describes how the stuttered speech impacts the various characteristics like formant transitions, pitch, VOT, etc. Section \ref{datasets} presents stuttering datasets that have been used in the ASIS. Section \ref{Statistical Approaches} describes various acoustic features like MFCC, Linear Predictive Cepstral Coefficients (LPCC), etc, that have been exploited in SD. This section also discusses all statistical machine learning methods, that have been used in various SD events. Section \ref{Deep Learning Approaches} describes, how deep learning can be used to model and detect various types of stuttering disfluencies. This section also provides some preliminary studies of DL in SD and discusses the use of self-supervised learning in SD. Section \ref{Challenges} discusses various challenges that the current ASIS systems are facing. Among them, a few are data scarcity, hand-engineered features, cross-domain, etc. This section also describes their possible solutions, that can be exploited to address the mentioned challenges, and, finally, the concluding remarks are provided in section \ref{Conclusion}.

\section{Characteristics of Stuttered Speech}
\label{acousticcharecteristics}
Most brain scan studies show that, during fluent speech and silent rest, there is no difference in cerebral activity between PWS and normal fluent speakers~\cite{ingham1996functional}. However, during stuttering, there is a dramatic change in cerebral activity. The right-hemisphere areas which are normally not active during normal speech become active, and the left-hemisphere areas, which are active during normal speech become less active~\cite{ingham1996functional,kehoe2006speech}. It has also been found that there is under-activity in the central auditory processing area. In 1997, a study by Kehoe \emph{et al}~\cite{kehoe2006speech}, suggests that adult PWS have an inability to integrate somatic and auditory processing. A brain study by Foundas \emph{et al}~\cite{foundas2001anomalous} found that PWS has rightward asymmetry in planum temporale (PT), i.e, their right PT is larger than their left PT, on the contrary, normal people's PT is larger in the left side of their brains. Studies based on motor data have been carried out about stuttering. The PWS also shows over-activity in the speech motor control area~\cite{smith2017stuttering, kehoe2006speech}, in particular in the \emph{left caudate nucleus} area. Conture \emph{et al} \cite{conture1977laryngeal,conture1985laryngeal} observe inappropriate vocal folds abductions and adductions which lead to anarchic openings and closure of the glottis. Concerning the supraglottic level, Wingate \emph{et al}~\cite{wingate1969stuttering} hypothesizes that stuttering is not a sound production problem but a coarticulation one. He theorizes that disfluencies occur during a fault line, which corresponds to the interval when muscular activity due to a sound that has been produced is going off and muscular movements for the following sound are going on. However in a recent study, Didirková and Hirsch~\cite{didirkova2020two} show, thanks to EMA data, that stuttering is not a coarticulation trouble. They found correct coarticulatory patterns in the fluent and stuttered utterances. Furthermore, another study based on articulatory observation, notes that stuttered disfluencies and non-pathological disfluencies do have common characteristics. However, stuttered disfluencies and non-pathological disfluencies produced by PWS present some particularities, mainly in terms of retention and anticipation, and the presence of spasmodic movements~\cite{didirkova2020articulatory}. PWS tend to develop strategies allowing them to avoid sounds or words which can result in disfluency; such strategies consist in using paraphrases or synonyms instead of the problematic segment \cite{ward2008stuttering}.

\par 
Concerning stuttering-like disfluencies, several types have been observed: repetitions, blocks, prolongations, interjections, etc, which are detailed in Table.~\ref{tab:stutter_types}. Some works try to link the locus of disfluencies and phonetic proprieties. Jayaram \emph{et al}~\cite{jayaram1983phonetic}, Blomgren \emph{et al}~\cite{blomgren2012speech}, and Didirkova \emph{et al}~\cite{didirkova2016parole} indicate that unvoiced consonants are more disfluent than their voiced counterparts. Furthermore, Blomgren \emph{et al}~\cite{blomgren2012speech} notices that disfluencies are more frequent at the beginning of an utterance or just after a silent pause. Moreover, Didirkova~\cite{didirkova2016parole} observes an important inter-individual variability concerning sounds and/or phonetic features which are the most disfluent.
\par 
Acoustic analysis has been carried out about stuttering, including speech rate, stop-gap duration, vowel(V)-consonant(C) transition duration, fricative duration, voice on-set time (VOT), CV transition duration, vowel duration, formants, glottis constriction, a sharp increase in articulatory power and closure length elongation before the speech segmented is released \cite{zebrowski1985acoustic}. 
\par 
Dehqa \emph{et al} \cite{dehqan2016formant} studied the second formant (F2) transitions of fluent segments of Persian speaking PWS. They concluded that the F2 frequency extent transitions are greater in stuttering speakers than in normal fluent ones. The PWS takes longer to reach vowel steady state, but the overall F2 formant slopes are similar for both stuttering speakers and normal ones \cite{dehqan2016formant}. The PWS generally exhibit slower speaking rates when compared to normal speakers. Several other studies have investigated the CV formant transitions in stuttered speech. Yaruss and Conture~\cite{yaruss1993f2} examined the F2 transitions of children who stutter on syllable repetitions and found no aberrant F2  transitions. However, Robb \emph{et al}~\cite{robb1998formant} analyzed the fluent speech segments of PWS and showed that F2 fluctuations are longer for voiced and voiceless stops than normal speakers. In a different study by Chang \emph{et al}~\cite{chang2002coarticulation}, where 3-5 year aged children were analyzed in the picture-naming task. The results showed that disfluent children produced smaller fluctuations of F2 transitions between the alveolar and bilabial place of articulations than did fluent children, and the overall degree of CV coarticulation is no different among stuttering and control groups. Subramanian \emph{et al}~\cite{subramanian2003second} analyzed the F2 frequency fluctuations of voiceless stops and revealed that near the onsets of CV,  the stuttering children exhibited smaller F2 changes than the normal speakers. Blomgren \emph{et al}~\cite{blomgren1998note} found that PWS and normal speaker in /CVt/ token exhibit no differences in the F1 (average). The stutters show significantly lower F2 in /Cit/ tokens than the control groups. The formant spacing for /i/ is significantly lower in PWS than fluent persons~\cite{blomgren1998note}. Hirsch \emph{et al}~\cite{hirsch:halshs-00716583} conducted a study by analyzing the first two formants (vowel space) in CV sequences between the stuttered and normal groups. At a normal speaking rate, the stuttering group shows a reduction in the vowel space, contrary to the fast speaking rate, where, the latter shows no noticeable deviation.

\par 
VOT is the duration of time between the release of a stop consonant and the beginning of vocal fold vibrations~\cite{guitar2013stuttering}. Healey and Raming~\cite{healey1986acoustic} showed that for voiceless stops, chronic stuttering exhibits longer VOT when compared with normal fluent persons. They showed that consonant and vowel duration were longer only in real-world phrases like \textit{take the shape}  in contrast with nonsense phrases like \textit{ipi saw ipi} \cite{healey1986acoustic}. Hillman and Gilbert~\cite{hillman1977voice} also found that the PWS reveals longer VOT for voiceless stops than for fluent persons. Adams~\cite{adams1987voice} found that not only voiceless stops exhibit longer VOT in PWS, but also, voiced stops display longer VOT than non-stuttering persons. No significant VOT differences have been found in control and PWS groups~\cite{watson1982comparison}. In another study by Jäncke~\cite{jancke1994variability}, the PWS show strong variability in repeated production of VOT for voiceless stops, however, there is no significant difference between the two groups. In a study carried out by De Nil and Brutten~\cite{de1991voice}, it shows that the stuttering children exhibit more variability in VOT than their counterparts. Celsete and de Oliveira martins-Reis~\cite{celeste2015impact} also found that the stuttering group shows higher VOT for voiceless stops. Brosch \emph{et al} \cite{brosch2002prognostic} examined the VOT in stuttering children. They found that the children with severe stuttering have higher values of VOT. Borden \emph{et al}~\cite{borden1985onset} examined the VOT of fluent utterances from PWS. Their study showed that the fluent utterances of PWS exhibit no statistical differences in VOT and are within the limits of normal groups. 
\par 
Fosnot and Jun~\cite{fosnot1999prosodic} examined the prosodic characteristics of PWS and fluent children. They found that the variability in pitch is greater in the stuttered group, but slightly differs from the normal group. In another study, it has been shown that the normal group and PWS show the same patterns in f0 deviation~\cite{ramig1981vocal}. The stuttering occurs less significantly in low-pitched conditions as compared to high-pitched condition~\cite{ramig1981vocal}

\section{Datasets for Stuttering Detection Research}
\label{datasets}
Data is an indispensable component of any DL model. DL saves feature engineering costs by automatically generating relevant features, however, requires substantial amounts of annotated data. Most stuttering identification studies so far are based on in-house datsets~\cite{howell2009university,rudzicz2012torgo, kourkounakis2020detecting, sep28k, ratner2018fluency} with limited speakers. In the stuttering domain, there is a lack of datasets and several stuttering datasets that have been collected so far are discussed below:

\paragraph{UCLASS} 
The most common concern in stuttering research is the lack of training data. University College Londons  Archive of  Stuttered  Speech (UCLASS) public dataset (although very small) \cite{howell2009university} is the most commonly used amongst the stuttering research community. The UCLASS comes in two releases from the UCL's Department of  Psychology and Language Sciences. This contains monologues, conversations, and readings with a total audio recording of 457. Some of these recordings contain transcriptions like orthographic, phonetic, and standard ones. Of these, orthographic are the ones that are best suitable for stutter labelling. The UCLASS\footnote{url:http://www.uclass.psychol.ucl.ac.uk/uclass1.htm} release 1 contains 139 monologue samples from 81 PWS, aged from five to 47 years. The male samples are 121 and the female samples are 18. 
 
\paragraph{LibriStutter} 
The availability of a small amount of labelled stuttered speech led to synthetic LibriStutter's creation \cite{kourkounakis2020detecting}. The LibriStutter consists of 50 speakers (23 males and 27 females), is approximately 20 hours and includes synthetic stutters for repetitions, prolongations, and interjections. For each spoken word, Kourkounakis \emph{et al}\cite{kourkounakis2020detecting} used Google Cloud Speech-to-Text (GCSTT) API to generate timestamp correspondingly. Random stuttering was inserted within the four-second window of each speech signal. 
\paragraph{TORGO} 
This was developed by a collaboration between departments of Speech Language Pathology Computer Science at the University of Toronto and the Holland-Bloorview Kids Rehab hospital \cite{rudzicz2012torgo}. This dataset comprises samples from seven persons, diagnosed with cerebralpalsy or amyotrophic lateral sclerosis including four males and three females aged between 16 to 50 years. In addition to this, it also contains samples from control speakers of the same age.
\paragraph{FluencyBank} 
This is a shared database for the study of fluency development which has been developed by Nan Bernstein Ratner (University of Maryland) and Brian MacWhinney (Carnegie Mellon University)~\cite{ratner2018fluency}. The platform proposes audio and video files with transcriptions of adults and children who stutter. The FluencyBank is an interview data of 32 PWS. 
\paragraph{SEP-28k}
The public stuttering datasets are too small to build well generalizable ASIS. So to address this, Colin et al~\cite{sep28k} recently curated a public version of stuttering events in podcasts (SEP-28k) dataset. This dataset contains total samples of 28,177. The SEP-28k dataset is the first publicly available annotated dataset with stuttering labels including (prolongations, repetitions, blocks, interjections, fluent(no disfluencies)) and non-disfluent labels including (natural pause, unintelligible, unsure, no speech, poor audio quality, and music) 
\paragraph{KSoF}
The Kassel State of Fluency (KSoF) is a German stuttered dataset collected recently and is comprised of 5597 three-second stuttered speech clips extracted from 214 recordings of 37 PWS with 28 males and 9 females~\cite{bayerl2022ksof}. This dataset contains a recording from three types of environments including telephonic conversation, read speech, and spontaneous speech. After collecting the stuttered recordings, the samples were downsampled to 16 kHz. This dataset also provides meta information in terms of gender, microphone type, and speaker identity. Like SEP-28k, the KSoF dataset also contains two types of labels: stuttering and non-stuttering labels. The stuttering labels contain core behaviors with blocks, prolongations, repetitions, interjections, and fluent segments, while non-stuttering labels include natural pause, unintelligible, unsure, no speech, poor audio quality, and music. 

\section{Automatic Stuttering Identification}
\subsection{Statistical Approaches}
\label{Statistical Approaches}
Stuttering identification is an interdisciplinary research problem in which a myriad number of research works (in terms of acoustic feature extraction and classification methods) are currently going on with a focus on developing automatic tools for its detection and identification. Most of the existing work detects and identifies stuttering either by language models~\cite{Zayats2016, chen} or by ASR systems~\cite{Alharbi2018, ALHARBI2020101052}, which first converts the audio signals into their corresponding textual
form, and then by the application of language models, detects or identifies stuttering. This section provides in detail a comprehensive review of the various acoustic-based feature extraction and machine learning stuttering identification techniques, that have been used in the literature.
\paragraph{Acoustic Features:} In the case of developing any speech recognition or stuttering identification system, representative feature extraction and selection is extremely an important task that affects the system performance. The first common step in the speech processing domain is feature extraction. With the help of various signal processing techniques, we aim to extract the features that compactly represent the speech signal and approximate the human auditory system's response \cite{huang2001spoken}. 
\par 
The various feature extraction methods that have been explored in the stuttering recognition systems are autocorrelation  function  and  envelope  parameters \cite{howell1995automatic}, duration, energy peaks, spectral of word based and part word based \cite{howell1997development1, howell1997development2, khara2018comparative}, age, sex, type of disfluency, frequency of disfluency, duration, physical concomitant, rate of speech, historical, attitudinal and behavioral scores, family history \cite{geetha2000classification}, duration and frequency of disfluent portions, speaking rate \cite{noth2000automatic}, frequency, $1^{st}$ to $3^{rd}$ formant frequencies and their amplitudes \cite{czyzewski2003intelligent, khara2018comparative}, spectral measure (fast Fourier transform (FFT) 512) \cite{szczurowska2014application, swietlicka2009artificial}, mel frequency cepstral coefficients (MFCC) \cite{chee2009overview, khara2018comparative, hariharan2012classification, esmaili2017automatic}, Linear Predictive Cepstral Coefficients (LPCCs) \cite{khara2018comparative, hariharan2012classification}, pitch, shimmer \cite{lopez2018analysis}, zero crossing rate (ZCR) \cite{khara2018comparative}, short time average magnitude, spectral spread \cite{khara2018comparative}, linear predictive coefficients (LPC), weighted linear prediction cepstral coefficients (WLPCC) \cite{hariharan2012classification},   maximum autocorrelation value (MACV) \cite{khara2018comparative}, linear prediction-Hilbert transform based MFCC (LH-MFCC) \cite{mahesha2017lp}, noise to harmonic ratio, shimmer harmonic to noise ratio , harmonicity, amplitude perturbation quotient, formants and  its  variants  (min,  max,  mean,  median,  mode, std),  spectrum centroid \cite{lopez2018analysis}, Kohonen's self-organizing Maps \cite{swietlicka2009artificial}, i-vectors \cite{ghonem2017classification}, perceptual linear predictive (PLP) \cite{esmaili2017automatic},  respiratory biosignals \cite{villegas2019novel}, and sample  entropy  feature \cite{hariharan2012speech}. With the recent developments in convolutional neural networks, the feature representation of stuttered speech is moving towards spectrogram representations from conventional MFCCs. One can easily discern the fluent and stuttered part of speech by analyzing the spectrograms as shown in \Cref{fig:spec}. Kourkounakis \emph{et al}~\cite{kourkounakis2020detecting} exploited the use of spectrograms (as a gray scale image) as a sole feature extractor for stutter recognition and thus makes it suitable for the convolutional neural networks.

\par 
Different speech parameterization methods have their benefits and drawbacks. Mahesha and Vinod~\cite{mahesha2013classification} compared LPC, LPCC, and MFCC for syllable repetition, word repetition, and prolongation and showed that LPCC-based multi-class SVM (92\% acc.) outperforms  LPCC (75\% acc) and MFCC(88\% acc) based SVM stutter recognition models.  Hariharan \emph{et al}~\cite{hariharan2012classification} discussed the effect of LPC, LPCC, and WLPCC features for stuttering (repetition and prolongation only) recognition events. They also discussed the effect of frame length and percentage of frame overlapping on stuttering recognition models. The authors conclude that the WLPCC feature-based stuttering recognition models outperforms LPC and LPCC. Fook {et al}~\cite{fook2013comparison} compared and analyzed the effect of LPC, LPCC, WLPCC, and PLP features on the repetition and prolongation type of disfluencies and it has been shown that the MFCC feature-based stuttering recognition models surpass the LPC, LPCC and WLPCC based ones. Arjun \emph{et al}~\cite{arjun2020automatic} used LPC and MFCCs as input features and concluded that MFCCs performs better than LPCs. Ai \emph{et al}~\cite{ai2012classification} performs a comparative study of LPCC and MFCC features in repetition and propagating stuttering and reports that LPCCs based ASIS outperforms MFFCs based ASIS slightly in varying frame length and frame overlapping. The optimal results of 94.51\%  and 92.55\% accuracy on 21 LPCC $\&$ 25 MFCC coefficients respectively have been reported \cite{ai2012classification}. This can be due to the possibility of LPCCs' potential in capturing the salient cues from stuttering  utterance~\cite{ai2012classification}. The use of spectrograms showed state-of-the-art performance in recognizing the stuttering events \cite{kourkounakis2020detecting}. The work by Kourkounakis \emph{et al} \cite{kourkounakis2020detecting} didn't focus on the blocks and multiple stuttering types if present in a speech segment.

 \setlength\LTleft{-2.6cm}
\fontsize{8}{11}\selectfont
{\begin{longtable}{ccccc} 
   \caption{Summary of several ASIS Systems in chronological order. (B: Block, F: Fluent, R: Repetition, P: Prolongation, In: Interjection, RAV: Respiratory air volume RAF: Respiratory air flow, MS: Modulation spectrum, SpR: Speech rythm, $a$: Audio modality, $b$: Audio and textual modality, $c$: Audio, textual and visual modality, $d$: Bio-respiratory signals )}\\
\toprule 
 Author and Year & Datasets & Features & Stutter Type & Model  \\
\midrule 
   Howell and Sackin~\cite{howell1995automatic}~(95) &  6 Speakers& EP, ACF-SC  &P, R & ANN~$^a$   \\
      \hline
        Howell \emph{et al}~\cite{howell1997development1,howell1997development2}~(97) & 12 Speakers & Energy peaks,  & NA & ANN~$^b$ \\
        &&Duration&&\\
        \hline
      Nöth \emph{et al}~\cite{noth2000automatic}~(00) & Northwind and Sun  & Disfluent Frequency, &NA &HMMs~$^a$ \\
        &37 Stutters, &Speaking rate, &&\\
        & 16 Non-Stutters &Duration&&\\
        \hline
        Geetha \emph{et al}~\cite{geetha2000classification}~(00) & 51 Stutters &Gender, Age&NA&ANNs~$^a$ \\
        & &Duration, Speech Rate&&\\
 
        \hline 
        Czyzewski \emph{et al}~\cite{czyzewski2003intelligent}~(03)&6-Normal, &Formants($1^{st}$ to $3^{rd}$),&P, R, SG&Rough Sets   \\
        &6-SG Samples&Amplitude&&ANNs~$^a$\\
         \hline
        Suszyński \emph{et al}~\cite{suszynski2015prolongation} (03)&NA&FFT&P&Fuzzy Logic~$^a$\\
        \hline
      Szczurowska \emph{et al}~\cite{szczurowska2014application}~(06)&8 PWS&FFT 512&B&MLP \\
        &&Spectral Measure&&SOM~$^a$\\
         \hline
        Wiśniewski \emph{et al}~\cite{wisniewski2007automatic}~(07)&30 samples&MFCCs&NA&HMMs~$^a$\\
 
        \hline
        Tan \emph{et al}~\cite{tan2007application}~(07)& UTM Skudai &MFCCs &NA&HMMs~$^a$\\
        &10 Speakers (7M, 3F)&&&\\
         \hline
        Ravikumar \emph{et al}~\cite{ravikumar2008automatic}~(08)&10 PWS&MFCCs,&SR&Perceptron~$^a$\\
        && DTW for &&\\
        &&Score Matching&&\\
        \hline
        Świetlicka \emph{et al}~\cite{swietlicka2009artificial}~(09)&8 PWS (Aged 10-23)&FFT 512 &NA&Kohonen based ML~$^a$\\
         &4 Fluent (2M, 2F)&Spectral Measure&&Kohonen based RBF\\
         \hline
          Chee \emph{et al}~\cite{chee2009mfcc}~(09)&UCLASS&MFCCs&R, P&$k$-NN, LDA~$^a$\\
          \hline
          Chee \emph{et al}~\cite{chee2009automatic}~(2009)&UCLASS&LPCCs&R, P&$k$-NN, LDA~$^a$\\
          
          \hline
        Ravikumar \emph{et al}~\cite{ravikumar2009approach}~(09)&15 PWS&MFCCs, DTW for &SR&SVM~$^a$\\
        &&score matching&&\\
        \hline
        Yildirim and Narayanan \emph{et al}~\cite{yildirim2009automatic}~(09)&10 CWS(Aged 4-6)&Duration, Pitch,  &R, FS, FP, RP, RP &NBC~$^c$\\
        &&Energy, Gestural,&&\\
        &&Linguistic&&\\
         \hline
        Pálfy and Pospíchal \emph{et al}~\cite{palfy2011recognition}~(11)&UCLASS&MFCCs&R&SVM(Linear Kernel)~$^a$\\
      &&&&SVM(RBF Kernel)\\
        \hline
        
        Mahesha and Vinod \emph{et al}~\cite{mahesha2013classification}~(13)&UCLASS&LPCC, MFCC&P, WR, SR&SVM~$^a$\\
        \hline
        Świetlicka \emph{et al}~\cite{swietlicka2013hierarchical}~(13)&19 PWS&FFT(512)&B, P, SR&Hierarchical ANN~$^a$\\ 
        \hline
        Oue \emph{et al}~\cite{oue2015automatic}~(15)&TORGO&MFCCs, LPCCs&R&DBN~$^a$\\
        \hline
        Mahesha and Vinod \emph{et al}~\cite{mahesha2017lp}~(17)&UCLASS&LH-MFCC&P, R, I&GMMs~$^a$\\
        \hline
        Esmaili \emph{et al}~\cite{esmaili2017automatic}~(17)&UCLASS&PLP&P&Correlation~$^a$\\
        \hline
    Esmaili \emph{et al}~\cite{esmaili2017automatic}~(17)&UCLASS&WPT with entropy&P, R&SVM~$^a$\\
      Esmaili \emph{et al}~ \cite{esmaili2017automatic}~(17)&Persian&WPT with entropy&P, R&SVM~$^a$\\
        \hline
        Ghonem \emph{et al}~\cite{ghonem2017classification}~(17)&UCLASS&I-Vectors&P, R, RP& $k$-NN, LDA~$^a$ \\ 
         \hline

        Santoso \emph{et al}~\cite{santoso2019classification}~(19)&UUDB, PASD&MS, SpR&NA&BiLSTM~$^a$ \\ 
        \hline
         Santoso \emph{et al}~\cite{santoso2019categorizing}~(19)&UUDB, PASD&MS, SpR &NA&BiLSTM + Attention~$^a$\\ 
        \hline
         Villegas \emph{et al}~\cite{villegas2019novel}~(2019)&69 Participants&Heart Rate, RAV, RAF &B&MLP~$^d$\\
        \hline
        Kourkounakis \emph{et al}~\cite{kourkounakis2020detecting}~(20) &UCLASS&Spectrograms&WR, SR, PR, I, P, Rv&ResNet $+$ BiLSTM~$^a$\\
        \hline 
        Kourkounakis \emph{et al}~\cite{kourkounakis2020fluentnet}~(20) &UCLASS, LibriStutter &Spectrograms&WR, SR, PR, I, P, Rv&FluentNet~$^a$\\
         \hline
        Sheikh \emph{et al}~\cite{sheikh:hal-03227223}~(21)&UCLASS&MFCCs&B, P, R, F&StutterNet~$^a$\\
        &128 PWS&&&\\
        \hline
        Sheikh \emph{et al}~\cite{stutternetmtl}~(22)&SEP-28k&MFCCs&R, P, B, In, F& MB \emph{StutterNet}\\
        \hline
        Sheikh \emph{et al}~\cite{stutternetmtl}~(22)&SEP-28k&MFCCs&R, P, B, In, F& MB \emph{StutterNet}\\
        \hline
        Sheikh \emph{et al}~\cite{sheikhw2v2}~(22)&SEP-28k&Speech embedddings&R, P, B, In, F& $k$-NN, NBC, ANNs\\
        \hline
        Bayerl \emph{et al}~\cite{bayerw2v2}~(22)&FluencyBank, KSoF&Speech embedddings&R, P, B, In, F& SVM\\
\bottomrule

\label{tab:sa}
\end{longtable}}
\normalsize

\paragraph{Machine Learning Classifiers:} Stuttering detection systems process and classify underlying stuttering embedded speech segments. Including traditional classifiers, many statistical machine learning techniques have been explored in the automatic detection of stuttering. However, the studies are empirical, so no generally accepted technique can be used. \Cref{tab:sa} lists chronologically the summary of stuttering classifiers including datasets, features, modality, and stuttering type.
\par 
In ASIS, typically classification algorithms are used. A classification algorithm approximates the input \textit{X} and maps it to output \textit{Y} by a learning procedure, which is then used to infer the class of the new instance. The learning classifier requires annotated data for training which discerns the samples into their corresponding labels/classes. Once the training is finished, the performance of the classifier is evaluated on the remaining test data.
\par 
The traditional classifiers that explore stuttering identification include support vector machines (SVM), hidden Markov models (HMM), perceptrons, multi-layer perceptrons (MLP), Gaussian mixture models (GMM), k-nearest neighbor ($k$-NN), naive Bayes classifier (NBC), rough sets, Kohonen maps (self-organizing maps (SOM)), linear discriminant analysis (LDA), artificial neural networks (ANN)
\paragraph{Hidden Markov Models} HMMs lie at the heart of all contemporary speech recognition systems and have been successfully extended to disfluency classification systems. A simple and effective framework is provided by HMMs for modeling temporal sequences. 
Wiśniewski \emph{et al}~\cite{wisniewski2007automatic} used Euclidean distance as a codebook based on 20 MFCCs with HMMs. They reported an average recognition rate of 70\% for two stuttering classes including blocks and prolongation with deleted silence and 60 frames of window length. Tan \emph{et al}~\cite{tan2007application} used 12 MFCC features with HMMs. The average recognition rate is 93\% \cite{tan2007application}. This tool recognizes only normal and stutter utterances and is not classifying different types of disfluencies. In 2000, Nöth \emph{et al}~\cite{noth2000automatic} used speech recognition system to evaluate the stuttering severity. This system can perform statistical counting and classification of three different types of disfluencies including repetition, pauses, and phoneme duration. Frequency of disfluent segments, speaking rate, and disfluent durations are the measurable factors used to evaluate the stuttering severity during therapy sessions \cite{chee2009automatic}   
 
\paragraph{Support Vector Machines} SVMs gained substantial attention, and have been widely used in the area of speech domain. SVM is a linear classifier that separates the data samples into their corresponding classes by creating a line or hyperplane. Mahesha and Vinod~\cite{mahesha2013classification} used multiclass SVM to classify three stuttering disfluencies including prolongations, word repetitions, and syllable repetitions. In this study, the different acoustic features including  12 LPC, LPCC, and MFCCs are used. 75\% average accuracy is obtained for LPC based SVM, whereas LPCC based SVM is 92\% and for MFCCs based SVM is 88\% \cite{mahesha2013classification}. Ravikumar \emph{et al}~\cite{ravikumar2009approach} used SVM to classify one disfluency type which is syllable repetitions. The features used are MFCCs and DTW for score matching. Average accuracy of 94.35\% is obtained on syllable repetitions. Pálfy and Pospíchal~\cite{palfy2011recognition} used SVM with two different kernel functions including linear and radial basis function (RBF). In this case study, they used 16 audio samples from UCLASS \cite{howell2009university} with eight males and eight females. 22 MFCC acoustic features with hamming window (25~ms) with an overlap of 10~ms are used in this case study \cite{palfy2011recognition}. 96.4\% is the best recognition rate that has been reported with SVM when RBF is used as a kernel function \cite{palfy2011recognition}. With linear kernel-based SVM, recognition rate is 98\% \cite{palfy2011recognition}. 
Esmaili \emph{et al}~\cite{esmaili2017automatic} used PLP features with a hamming window of 30~ms and an overlap of 20~ms to detect the prolongation type of stuttering based on a correlation similarity measure between successive frames. 99\% and 97.1\% is the best accuracy that has been reported on UCLASS and Persian datasets respectively \cite{esmaili2017automatic}. In the same study they also evaluated the WPT+entorpy feature-based SVM on UCLASS and Persian stuttering datasets with 99\% and 93.5\% accuracies respectively \cite{esmaili2017automatic}.

\paragraph{Artificial Neural Networks (ANNs)} They consist of several connected computing neurons that loosely model the biological neurons \cite{goodfellow2016deep}. Like the synapses in biological neurons, each neuron can transmit a signal to other neurons via connections. A neuron receives a signal, processes it, and can transmit a signal to other connected neurons. The connections have weights associated with them which adjusts the learning procedure \cite{goodfellow2016deep}. 
ANNs are trained by processing examples that map an input to its corresponding result by forming probability-weighted associations between the two. The training is conducted with the help of backpropagation by optimizing the loss function by computing the error difference between the predicted output and its corresponding ground truth. Continuous weight adaptations will cause the ANNs to produce a similar output as the ground truth. After an adequate number of weight adjustments, the training can be terminated once the optimization criteria are reached \cite{goodfellow2016deep}. ANNs are essential tools both in speech and speaker recognition. In recent times, ANNs play important roles in identifying and classifying the stuttering speech. Howell \emph{et al}~\cite{howell1995automatic} used two ANNs for repetition and prolongation recognition. The neural net is trained with 20 ACF, 19 vocoder coefficients of 10~ms frame length, and also with 20 frames of envelope coefficients. The networks are trained in just two minutes of speech. The best accuracies of 82\%  and 77\%  are obtained for prolongations and repetitions when envelope parameters are used as input features to ANNs \cite{howell1995automatic}. ACF-SC-based ANNs give the best accuracy of 79\% and 71\% for prolongations and repetitions respectively \cite{howell1995automatic}. P. Howell \emph{et al}~\cite{howell1997development1, howell1997development2} designed a two-stage recognizer for the detection of two types of disfluencies including prolongation and repetitions. The speech is segmented into linguistic units and then classified into its constituent category. The network is trained with the input features including duration and energy peaks on a dataset of 12 speakers \cite{howell1997development1, howell1997development2}. The average accuracy on prolongations and repetitions obtained in this case study is 78.01\% \cite{howell1997development1,howell1997development2}. Geetha~\emph{et al}~\cite{geetha2000classification} studied ANNs on 51 speakers to differentiate between stuttering children and normal disfluent children based on the features including disfluent type, rate of speech, disfluency duration, gender, age, family history, and behavioral score. They reported a classification accuracy of 92\% \cite{geetha2000classification}. Szczurowska \emph{et al}~\cite{szczurowska2014application} used Kohonen based MLP to differentiate between non-fluent and fluent utterances. 76.67\% accuracy has been reported on blocks and stopped consonant repetition disfluency types~\cite{szczurowska2014application}. The Kohonen or self-organizing maps (SOM) are used first to reduce the feature dimensions of FFT 512 (with 21 digital 1/3-octave filters and a frame length of 23~ms) input features, which later act as an input to the MLP classifier. The model is trained on eight PWS \cite{szczurowska2014application}. Ravikumar \emph{et al}~\cite{ravikumar2008automatic} proposed an automatic method by training a perceptron classifier for syllable repetition type of disfluency on 10 PWS with 12 MFCCs and DTW as the feature extraction methods. The best accuracy obtained for syllable repetition is 83\% \cite{ravikumar2008automatic}. In 2003, Czyzewski \emph{et al}~\cite{czyzewski2003intelligent} addressed the stuttering problem with the help of stop-gaps detection, identification of syllable repetitions, detecting vowel prolongations. They applied ANNs and rough sets to recognize the stuttering utterances on the dataset of six fluent and six stop-gap-based speech samples \cite{czyzewski2003intelligent}. They reported that the average prediction accuracy of ANNs is 73.25\% and rough-sets yielded average accuracies of 96.67\%, 90.00\%, 91.67\% on prolongations, repetitions, and stop-gaps respectively \cite{czyzewski2003intelligent}. Suszyński \emph{et al}~\cite{suszynski2015prolongation} proposed a fuzzy logic-based model for the detection and duration of prolongation type of disfluency. They used a sound blaster card with a sampling frequency of 22~kHz. 21 1/3 octave frequency bands with $A$ filter and FFT features are used with the hamming window of 20~ms. The features representing the prolongations are described by the use of fuzzy sets. Only the disfluent fricatives and nasals are considered in this study \cite{suszynski2015prolongation}. Świetlicka \emph{et al}~\cite{swietlicka2009artificial} proposed an automatic recognition of prolongation type of stuttering by proposing Kohonen-based MLP and RBF. From a dataset of eight PWS and four fluent speakers, 118 (59 disfluent, 59 fluent), 118 total speech samples are recorded for the analysis. 21 1/3 octave filters with frequencies ranging from 100~Hz to 10000~Hz are used to parametrize the speech samples \cite{swietlicka2009artificial}. The parametrized speech samples are used as input features to the Kohonen network that is expected to model the speech perception process. Thus, Kohonen is used to reduce the input dimensionality to extract salient features. These salient features are then fed to the MLP and RBF classifiers that are expected to model the cerebral processes, responsible for speech classification and recognition \cite{swietlicka2009artificial}. The method yielded a classification accuracy of 92\% for Kohonen-based MLP and 91\% for Kohonen-based RBF \cite{swietlicka2009artificial}.
\par 
Villegas \emph{et al}~\cite{villegas2019novel} introduced a respiratory bio-signals based stuttering classification method. They used respiratory patterns (air volume) and pulse rate as input features to MLP. The dataset, developed at the Pontifical Catholic University of Peru consists of 68 Latin American Spanish-speaking participants with 27 PWS (aged 18-27 with a mean of 24.4$\pm$5 years), 33 normal (aged 21-30 with a mean of 24.3$\pm$2.3 years). The stuttering type studied in this research work is blocks with an accuracy of 82.6\% \cite{villegas2019novel}. 
\par 
In 2013, Mahesha and Vinod~\cite{mahesha2017lp} introduced a new Linear prediction-Hilbert transform-based MFCC (LH-MFCC) human perception feature extraction technique to capture the temporal, instantaneous amplitude, and frequency characteristics of speech. The study compares the MFCC and LH-MFFC features for three types of disfluencies including prolongation, repetition, and interjection in combination with 64 Gaussian mixture model (GMM) components and reported a gain of 1.79\% in average accuracy \cite{mahesha2017lp} with LH-MFCCs. The proposed LH-MFCC improves discriminatory ability in all classification experiments \cite{mahesha2017lp}.
\paragraph{$k$-Nearest Neighbor and Linear Discriminant Analysis} $k$-NN, proposed by Thomas Cover is a non-parametric model that can be used for both classification and regression. In $k$-NN classification, the output is described by the class membership and a sample is classified by the contribution of its neighbors. The sample is assigned to the class which is most common among its \textit{k (k$\geq$0)} neighbors. This method relies on the distance metric for classification \cite{murphy2012machine}
\par 
LDA also called normal discriminant analysis (NDA), or discriminant function analysis is a technique used in statistics and machine learning, to find a linear combination of features that separates two or more classes of samples~\cite{murphy2012machine}. 
\par 
Chee \emph{et al}~\cite{chee2009mfcc} presented an MFCC feature based $k$-NN and LDA classification models for repetition and prolongation types of disfluencies. The proposed models reports the best average accuracies of 90.91\%  for $k$-NN (with $k$=1) and 90.91\% for LDA \cite{chee2009mfcc} on UCLASS \cite{howell2009university} dataset. In 2009, Chee \emph{et al}~\cite{chee2009automatic} studied the effectiveness of LPCC features in prolongation and repetition detection with $k$-NN and LDA classfiers. The work achieved an average accuracy of 87.5\% and the best average accuracy of 89.77\% for LDA and $k$-NN respectively on the UCLASS \cite{howell2009university} dataset. In 2017, Ghonem \emph{et al}~\cite{ghonem2017classification} introduced an I-vector (commonly used in speaker verification) feature based stuttering classification  with $k$-NN and LDA methods. The technique reported an average accuracy of 52.9\%  among normal, repetition, prolongation and repetition-prolongation\footnote{repetition and prolongation disfluencies appearing at the same time} stuttering events \cite{ghonem2017classification}. This is the first technique so far that has taken two disfluencies (occurring at same time) into consideration.
 \par 
 In 2009, Yildrim and Narayanan \emph{et al}~\cite{yildirim2009automatic} proposed the first multi-modal disfluency boundaries detection model in spontaneous speech based on audio and visual modalities.  The dataset used in this study was collected using the Wizard of Oz (WoZ) tool. Audio recordings of high quality were collected using a desktop microphone at 44.1~kHz. Two Sony TRV330 digital cameras, one focused from the front and the other capturing the child and the computer screen from the side were also used \cite{yildirim2009automatic}. Three different classifiers including $k$-NN, NBC, and logistic model trees have been utilized to evaluate the effectiveness of multi-modal features on the collected dataset \cite{yildirim2009automatic}. The stuttering types included in this case study are repetition, repair, a false start, and filled pauses \cite{yildirim2009automatic}. In this work, the combination of three different modality-based features including prosodic (duration, pitch, and energy), lexical (hidden event posteriors), and gestural (optical flow) features were studied at feature level and decision level integration  \cite{yildirim2009automatic}. The work achieved the best accuracy for NBC among the three classifiers \cite{yildirim2009automatic} and reports an accuracy of 80.5\% and 82.1\% at feature level integration and decision level feature integration respectively \cite{yildirim2009automatic}. 
\par 
In 2005, Oue \emph{et al}~\cite{oue2015automatic} introduced deep belief network for the automatic detection of repetitions, non-speech disfluencies. 45 MFCC and 14 LPCC features from TORGO  dataset \cite{rudzicz2012torgo} have been used in this case study for the detection of disfluencies \cite{rudzicz2012torgo}. The experimental results obtained showed that MFCCs and LPCCs produce similar detection accuracies of approximately 86\% for repetitions and 84\% for non-speech disfluencies \cite{oue2015automatic}.
\par 
The majority of statistical machine learning ASIS systems detailed above mostly focused only on either $prolongation$ or $repetition$ types of disfluencies with the most widely used features as MFCCs. Among the statistical techniques mentioned above, SVMs are the most widely used classifier in stuttering detection and identification.

\subsection{Deep Learning Approaches}
\label{Deep Learning Approaches}

The majority of the state-of-the-art deep learning techniques combine several non-linear hidden layers as it can reach hundreds of layers as well, while traditional ANNs consist of only one or two hidden layers. With the advancement in deep learning technology, the improvement in the speech domain surpasses the traditional machine learning algorithms, and hence the research in the speech domain shifts towards the deep learning-based framework, and SD is no exception. The salient advantage of these deep networks is automatic feature selection and extraction which avoids the cumbersome and tedious work of manual feature engineering step. The goal of these deep architecture classifiers is to approximate a mapping function $f$ with $\textbf{y} = f(\textbf{X};\theta)$ from input samples \textbf{X} to target labels $\textbf{y}$ by adjusting its parameters $\theta$. The most common deep learning architectures used in the ASIS research domain are convolutional neural networks and recurrent neural
networks.
\paragraph{Recurrent Neural Networks (RNNs)} 
RNNs belong to a family of deep neural architectures where connections between neurons/nodes form a directed graph along a temporal sequence, thus allowing it to show temporal dynamic behavior. RNNs consist of an internal state (memory) that is used to process variable-length input sequences. This structure makes RNNs good for modeling sequential tasks like time series, connected handwriting, video, or speech recognition \cite{goodfellow2016deep}. The other networks process inputs which are independent of each other, but in RNNs, inputs are related to each other.

 \par 
Long short-term memory networks (LSTMs) introduced by Hochreiter and Schmidhuber \emph{et al}~\cite{lstmorig}, is a special type of RNN, capable of capturing the long-term dependencies in the temporal sequence. 

 \par 
 In 2019, Santoso \emph{et al}~\cite{santoso2019categorizing} proposed modulation spectrum feature-based BiLSTM (Bidirectional LSTM) to detect the causes of errors in speech recognition systems. The method is tested on the Japanese dataset of 20 speakers with 10 males and 10 females \cite{santoso2019categorizing}. The experiment used a 640-dimensional modulation spectrum feature vector with a block length of 320~ms \cite{santoso2019categorizing}. The method achieved an F1-score of 0.381 for successfully detecting the stuttering events in the speech \cite{santoso2019categorizing}. The proposed model used the overall utterance for the stuttering error detection, however, recognition errors arise only from a small part of the full utterance. In order to address this issue, Santoso \emph{et al}~\cite{santoso2019classification} introduced attention based BiLSTM classifier for stuttering event detection. The best F1-score of 0.691 is attained by taking the block length of 32~ms \cite{santoso2019classification}. 
 \paragraph{Convolutional Neural Networks (CNN)}  
CNNs are a special type of neural nets that work with grid-structured data like images, audio spectrograms, video frames, etc. A CNN consists of several layers in the pipeline: convolution, pooling, and fully-connected layers. With the help of several feature maps, CNNs are successful in capturing the spatial and temporal dependencies from the input data. 
\par 
The convolution layer, a core component of the CNNs, is comprised of a set of learnable parametric kernels (filters) that transforms an input image into several small receptive fields \cite{goodfellow2016deep}. In the forward pass, a dot product is performed between the entries of an input image and filter resulting in an activation map of that filter \cite{goodfellow2016deep}. This dot product is also known as convolution operation, defined by the following equation: 

\begin{equation}
    feature \hspace{0.1cm} map = y[i,j] = input \circledast kernel = \sum \sum X[i-m, j-n].h[m,n]
\end{equation}
where $i,j$ indices related to image and $m,n$ are concerned with the kernel  , $X$ represents the audio spectrogram or image matrix which is to be convolved with the filter $h$.
\par 
Due to parameter sharing of the convolutional operation, divergent feature or activation maps can be extracted, thus making the CNNs translation invariance architectures \cite{goodfellow2016deep}. Pooling, a down-sampling dimensionality reduction layer partitions the input matrix into a set of translational invariant non-overlapping combinations of features. There are many methods to implement pooling operation, the most common among which is $average$ pooling, which computes the average value from each sub-region of the feature maps \cite{goodfellow2016deep}. Fully connected (FC) layers, a global operation unlike convolution and pooling, usually used at the end of the network, connects every neuron in one layer to every neuron in another layer \cite{goodfellow2016deep}. The FC layer takes the non-linear combination of selected features, which is later used for downstream tasks like classification \cite{goodfellow2016deep}. 
\par
\begin{figure}[h]
    \centering
    \includegraphics[scale=0.5]{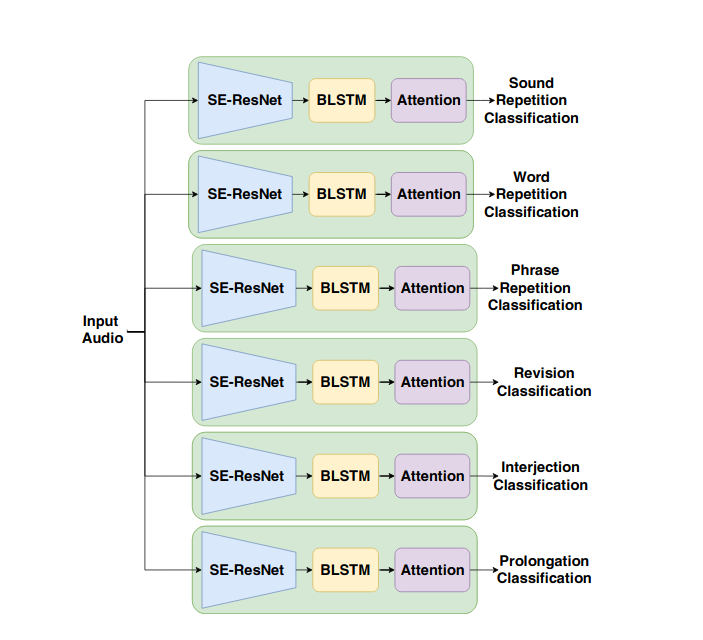}
    \caption{FluentNet model for stuttering classification (reproduced with permission taken from authors)~\cite{kourkounakis2020fluentnet}.}
    \label{fig:fluentnet}
\end{figure}

 
As discussed in \Cref{Statistical Approaches} that most of the existing stuttering identification work either depends on language models or ASR systems. This procedure of stuttering identification seems a subsidiary computational step and could also be a potential source of error. To address this, Kourkounakis \emph{et al}~\cite{kourkounakis2020detecting} proposed a CNN-based model to learn stutter-related features. They formulated SD as a binary classification problem, where they used the same architecture for identifying different types of stuttering. They used residual-based CNN and BiLSTM (for temporal aspect) to capture the disfluency-specific features from the spectrograms \cite{kourkounakis2020detecting}, that are the sole input features used in this study. The model is trained with batch norm and ReLU activation function \cite{kourkounakis2020detecting}. Each BiLSTM layer is followed by a dropout rate of 0.2 and 0.4 respectively \cite{kourkounakis2020detecting}. The proposed model reported an average accuracy of 91.15\% and an average miss rate of 10.03\%  (surpasses the state-of-the-art by almost 27\%) on six different types of stuttering: revision, prolongation, interjection, phrase repetition, word repetition, and sound repetition \cite{kourkounakis2020detecting}. Kourkounakis \emph{et al}~\cite{kourkounakis2020fluentnet} proposed a FluentNet as shown in \Cref{fig:fluentnet}. that combines squeeze-and-excitation residual network (SE-ResNet) with BiLSTM networks, where SE-ResNet (eight blocks) is used to learn the stutter-specific spectral frame-level representations. Each audio speech is first segmented into four-second audio clips, then acoustic features (spectrograms) are extracted, which are fed to SE-ResNet to capture stutter-specific spectral features, followed by a global attention-based two-layered BiLSTM (512 units) network, that helps in capturing effective temporal relationships \cite{kourkounakis2020fluentnet}. The model is trained using a root mean square propagation (RMSProp) optimizer on a binary cross entropy loss function with a dropout of $0.2$ and a learning rate of $10^{-4}$. 

\par 
In order to tackle the issue of stuttered speech data scarcity, they developed a synthetic stuttered speech dataset (LibriStutter) from a fluent LibriSpeech datatset \cite{kourkounakis2020fluentnet}. The proposed FluentNet model reports an average accuracy of 91.75\%  and 86.7\%  on UCLASS and LibriStutter datasets respectively. Six different disfluency types are considered in this experimental study including phoneme repetition, word repetition, phrase repetition, interjection, prolongation, and revisions \cite{kourkounakis2020fluentnet}. 
\par
\begin{figure}
    \centering
    \includegraphics[scale=0.5]{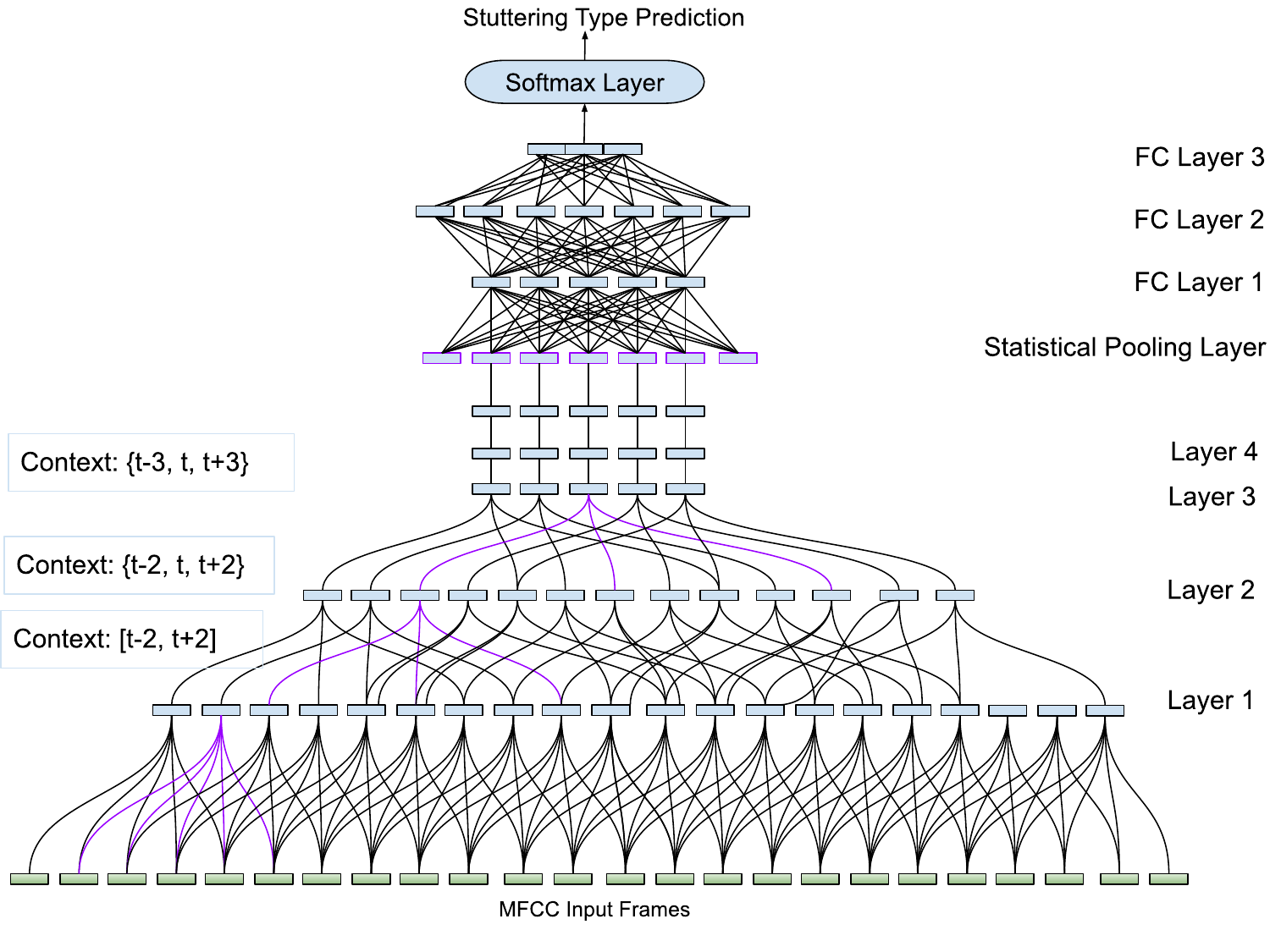}
    \caption{StutterNet (reproduced with permission taken from authors)~\cite{sheikh:hal-03227223}.}
    \label{fig:stutternet}
\end{figure}

 
The stuttering identification methods discussed above consider only a small subset of disfluent speakers in their experimental studies, so it can not be said with certainty that the discussed models, which performed very well on small speakers can also generalize to a large set of stuttered speakers. In order to evaluate this, Sheikh \emph{et al.} recently proposed a \textit{StutterNet} \cite{sheikh:hal-03227223}, a time delay neural network based SD method shown in \Cref{fig:stutternet}. They addressed this problem by formulating it as a multi-class classification problem. Only the core behaviors (blocks, repetition, and prolongation) and fluent segments of the speech were considered in this case study.  128 speakers from the UCLASS dataset were used in this case study, thus making it the first experimental study to be evaluated on a large set of disfluent speakers. Each audio sample is initially divided into four-second audio segments, then acoustic features (MFCCs) are extracted, which are then fed to the \textit{StutterNet}. The features are generated after every 10~ms on a 20~ms
window for each 4-sec audio sample. On this larger set of disfluent speakers, we compared this study with the ResNet+BiLSTM \cite{kourkounakis2020detecting} based ASIS system and reported an overall average accuracy of 50.79\%  and Mathews correlation coefficient
(MCC) of 0.23, in comparison to the ResNet+BiLSTM-based system comprising 46.10\%  overall average accuracy and 0.21 MCC. The comparative results are shown in \Cref{tab:Accuracy}.


\begin{table}
    \caption{Results in accuracies and MCC(B: Block, F: Fluent, R: Repetition, P: Prolongation, In: Interjection)}
    \scalebox{0.9}{\begin{tabular}{*{8}{c}}

    \toprule
    \multicolumn{1}{c}{Method}&\multicolumn{5}{c}{Accuracy}&\multicolumn{1}{c}{Tot. Acc.}&\multicolumn{1}{c}{MCC.}\\
    \midrule
    &R&P&B&F&In&&\\
    \midrule
        \multicolumn{8}{c}{\textbf{Dataset:~UCLASS}} \\
\midrule
    Resnet+BiLSTM~\cite{kourkounakis2020detecting}&
    20.39&	\textbf{23.17}&	\textbf{53.33}&	55.00&NA	&46.10&	0.20\\
    \midrule
    \textit{ \textbf{StutterNet}  (Baseline)}&
    \textbf{27.88}&	17.13&	42.43&	66.63&NA	&49.26&	0.21\\
    \textbf{\emph{StutterNet} (Optimized)}&
    23.98&	12.96&	47.14&	\textbf{69.69}&NA	&\textbf{50.79}&	\textbf{0.23}\\
    \midrule
    \multicolumn{8}{c}{\textbf{Dataset:~SEP-28k}} \\
    \midrule
    \emph{StutterNet}~\cite{sheikh:hal-03227223}&21.99&27.78&1.98&88.18&49.99&60.33&NA\\
    MB \emph{StutterNet}~\cite{stutternetmtl}&28.70&37.89&9.58&74.43&57.65&57.04&NA\\
    Sheikh \emph{et al}~\cite{sheikhw2v2}&\textbf{46.79}&\textbf{40.79}&\textbf{23.86}&84.32&\textbf{69.54}&\textbf{68.35}&NA\\
 \bottomrule
    \end{tabular}}
     
        \label{tab:Accuracy}
\end{table}

\par 
Among the DL-based ASIS systems described above in detail, for a small set of disfluent speakers, the FluentNet classifier proposed by Kourkounakis \emph{et al}~\cite{kourkounakis2020fluentnet} and the spectrogram feature representations of stuttered speech are the most effective, that gives promising classification results on disfluency identification. However, for a large set of stuttered speakers, \textit{StutterNet} is the most effective one.
\par
\begin{figure}
    \centering
    \includegraphics[scale=0.4]{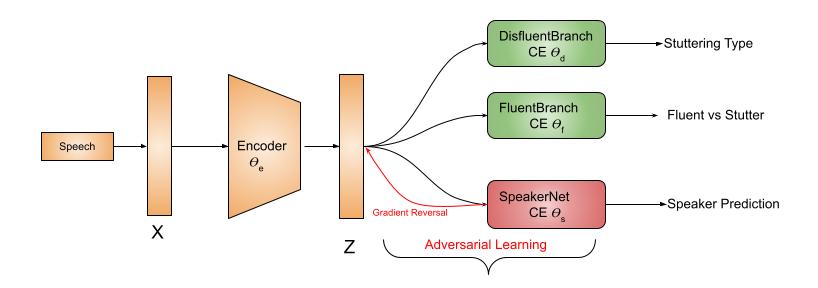}
    \caption{Stuttering detection using adversarial multi-task learning with red curve indicating gradient reversal, CE: Cross entropy, $\theta$'s are the parameters of each module (reproduced with permission taken
from authors)~\cite{stutternetmtl}.}
    \label{fig:advmtl}
\end{figure}

 
In a recent study by Lea \emph{et al}~\cite{sep28k}, they curated a large stuttering dataset named \emph{SEP$-$28k} and employed the ConvLSTM model to detect various types of stuttering. In addition to 40~MFCC input features, the model also takes the pitch and articulatory features as an input and reports a weighted accuracy of 83.6, F1 of 83.6 on the \emph{FluencyBank}. On \emph{SEP-28k}, they reported F1-scores of 55.9, 68.5, 63.2, 60.4, and 71.3 in the block, prolongation, sound repetition, and word repetition respectively. They also evaluated their proposed model on 41~dim phoneme probabilities extracted from pre-trained time-depth separable CNN on LibriSpeech and reports an F1 score of 74.8 on \emph{FluenyBank}. The model was trained with a mini-batch size of 256 and a cross entropy loss function. In another study by Melanie \emph{et al.}~\cite{melanie}, where they mix the SEP-28k, UCLASS, and FluencyBank datasets and used a phoneme feature-based BiLSTM model for SD.  SD systems are further impacted by source diversity, whether it be in the form of a speaker, gender, meta-data information (such as podcast ids), accent, linguistic content, or overall acoustic settings. Exploiting this fact, Sheikh \emph{et al}.~\cite{stutternetmtl} used podcast ids and proposed multitasking and adversarial-based robust SD via a multi-branched training scheme as shown in Fig.~\ref{fig:advmtl}. The model tries to learn stutter representations from a stutter speech utterance that are podcast invariant but at the same time are stutter discriminative. They found that the blocks are the hardest to detect in the SEP-28k dataset. In addition including fluent speech of PWS in training greatly affects its recognition performance~\cite{stutternetmtl, sheikhw2v2}. The application of advanced DL architectures for SD is constrained by the limited size and high cost of the voice pathology datasets. To this, Bayerl \emph{et al}.~\cite{bayerw2v2} used the Wav2Vec2.0 model pre-trained in a self-supervised fashion as a feature extractor to demonstrate its effectiveness in SD. They fine-tuned the Wav2Vec2.0 model on FluencyBank (English)~\cite{sep28k} and KSoF (German)~\cite{bayerl2022ksof} datasets and used SVM as a downstream binary classifier and reported an average F1 score of 53.8\% and 46.67\% on the KSoF and FluencyBank datasets respectively.
Similar research was conducted by Sheikh \emph{et al.}\cite{sheikhw2v2}, where they investigated the impact of speaker embeddings extracted from a pre-trained emphasized channel attention, propagation, and aggregation-time-delay neural network (ECAPA-TDNN) in SD in addition to speech embeddings from Wav2Vce2.0. They showed that the middle layers from Wav2Vec2.0 are good at capturing stuttering-related information. They also showed that fusing information from Wav2Vec2.0 and ECAPA-TDNN pre-trained can further improve the detection accuracy of stuttering types. In addition, they concatenated the information from several layers of Wav2Vec2.0 and reported the state-of-the-results (68.35\% overall accuracy) in SD on the SEP-28k dataset.
\par 
%

\section{Challenges \& Future Directions}
\label{Challenges}
This section describes various challenges faced by ASIS systems and their possible solutions, which can be explored in the field of stuttering research. Although there have been several developments in the automatic identification of stuttering, there are still several impediments that need to be addressed for robust and effective identification of stuttering.
\subsection{Dataset Collection}
 One of the most common barriers that need to be addressed is the issue of scarcity of data on stuttering identification research. There are only a few natural stuttered datasets as discussed in \Cref{datasets}. Usually, medical data collection is expensive and very demanding, and stuttering is no exception. Thus, having ample speaker and sentence variation adds more complexity to the stuttering domain. In order to make a fine analysis across several speakers, it is appropriate to have the same content (same list of sentences). Unfortunately, in practice, when a PWS is asked to read a list of sentences, the disfluency effects are greatly reduced. For this reason, more spontaneous speech is used in the hope to induce disfluencies. Moreover, depending on the speaker, the presence of disfluency in a recording is more or less important for several reasons: emotional state, speaking in public or alone, spontaneous or read speech, etc. This makes the collection of a corpus even more difficult and its size from one speaker to another can be variable if one aims at having a comparable number of examples of disfluencies. Moreover, it is extremely difficult, if not impossible, to collect a corpus that contains the same number of examples of each type of disfluency. It is even more challenging to achieve high variability in gender, language, and dialect. It should be noted that the recording of spontaneous speech must be well controlled to comply with the legislation. Due to the sensitivity of medical data and privacy concerns, it can not be applied at a large scale. Currently, we are not dealing with anonymization, as the voice could identify the speaker, but a minimum effort in this direction is required.
\par 
In order to identify stuttering using deep learning models, the data must be properly labelled. Different background noises can corrupt the stuttered speech data. Likewise, the noise of recording equipment can also degrade the speech signal. Noise injection techniques~\cite{yin2015noisy} can be exploited to learn reliable stutter-specific features from the noisy corrupted data. DL models like denoising auto encoders, imputation auto encoders \cite{latif2020deep} can also be utilized to learn robust stutter-specific features from corrupted data. Training and testing data distribution mismatch is a significant challenge for ASIS systems to be robust to noise. 

\par 
Since stuttering datasets are scarce, we can attempt to solve this problem by enlarging the training data size and its diversity by generative models. Deep generative models such as generative adversarial networks can be utilized for data augmentation~\cite{qian2019data} to generate more stuttered speech samples with the aim of improving the stuttering identification systems.

\subsection{Data Annotation Issue}
It is no doubt that DL has led to enormous advancement in the speech domain, nonetheless, it demands a large amount of labelled data, and also, the dataset bias has plagued current ASIS methods. Annotating the stuttered speech requires expert speech pathologists/therapists, thus is expensive and laborious. Unsupervised learning~\cite{goodfellow2016deep} can be used to exploit unlabelled data from different distributions, and these learned representations can, later on, be used in various (limited data, e.g., stuttering) downstream classification tasks. Unsupervised learning such as contrastive predictive coding~\cite{oord2018representation} enables to capture the underlying innate structure/pattern(s) in the data distribution. In the context of the stuttered speech, it can capitalize on the unlabelled data to create understandings and learn good stutter-specific feature representations, which later on, can be used to enhance the performance of ASIS systems in a supervised fashion. Semi-supervised learning~\cite{zhu2009introduction} can also be exploited to solve this problem by employing unlabelled data, in conjunction with the annotated data to develop better classification models. Due to the unavailability of the annotated and limited size of stuttering data, it becomes extremely difficult for the deep models to generalize. Self-supervision, where the main idea is to find a proxy or pretext task for the deep models to learn without any explicit annotations, but rather, the data's innate patterns provides the labels~\cite{shukla2020visually}, is a compelling approach to address this paucity of stuttered data by capturing the innate compositions of the disfluency data. By using the Wav2Vec2.0 and ECAPA-TDNN models as feature extractors, Sheikh \emph{et al}~\cite{sheikhw2v2} and Bayerl \emph{et al}~\cite{bayerw2v2} recently demonstrated the usefulness of leveraging features extracted from self-supervised pre-trained models that are trained on large datasets and reported the state-of-the art results in SD as shown in \Cref{tab:Accuracy}.

\subsection{Data Imbalance}
Stuttering datasets also suffer from data imbalance problems, i.e., the number of samples available for different disfluent categories is not uniform. It is mentioned that in stuttering, the repetitions are the most frequent ones followed by prolongations, and blocks~\cite{guitar2013stuttering}. However, in the UCLASS dataset, the block type is present in the majority followed by repetitions and prolongations. The model trained on this type of imbalanced dataset is biased towards the majority class. In order to address this, several techniques can be exploited, including resampling~\cite{chawla2002smote}, reweighting~\cite{cuiclassbalance} and metric learning~\cite{wang2018iterative}. Self-supervision as proposed recently by Yang et al. \cite{yang2020rethinking} can also be used to address the problem of labeling bias effect in learning on imbalanced disfluent data.

\subsection{Lack of Appropriate Acoustic Representation}
Another issue in the stuttering speech domain is the need for hand-engineered features, which approximate the human auditory system. MFCCs are the principal set of hand-engineered acoustic features that have been used mainly for stuttering identification tasks. The main drawback of this approach is that being manual is cumbersome and requires human knowledge. Over the past few years in the speech domain, the use of hand-engineered acoustic features is gradually changing and representation learning is acquiring recognition as an effective alternative to learn and capture task-specific features directly from raw speech signals, thus circumvents the hand-engineered feature extraction module from the pre-processing pipeline \cite{latif2020deep}.
Liu \emph{et al}~\cite{Xuechen9401593} recently proposed a learnable MFCCs for speaker verification. Sailor and Patil~\cite{sailor2016unsupervised} showed that an unsupervised deep auditory model can learn human auditory processing relevant features like filterbanks from raw speech. In addition, Millet and Zeghidour~\cite{millet2019learning} used raw signals to learn the filterbanks for dysarthria detection.
This could be exploited to learn and capture the stuttered-specific features directly from the raw speech signal, which later on can be used for downstream tasks like classification, prediction, etc.  

\subsection{Domain Adaptation} 
Most of the existing ASIS techniques proposed so far are evaluated on a dataset of specific languages consisting of limited speakers. The existing ASIS techniques depend on a probabilistic model to capture domain-specific factors so that any alteration in the input speech domain could have a significant impact (in terms of language or speakers) at the time of inference. It is yet to be explored that, how well an ASIS technique performs in a cross-domain or cross-language environment. There could be two possible scenarios for the cross-language issue: the first is when the model is trained with a specific-language data, but tested in other languages; the other scenario could be, during training, a disfluent person registered in one language, but evaluated in a different language at the test time. Learning stutter-specific features that are invariant to variabilities in language, speakers, recording conditions, etc., could improve the performance of ASIS systems. Domain adaptation techniques have been successfully applied in various speech tasks such as emotion recognition~\cite{daemotion}, speaker verification~\cite{Alam2018}, and ASR~\cite{samarakonASR}. However, in stuttering domain research, it has not been exploited yet. Several domain adaptation techniques such as~\cite{ganin2015unsupervised, wang} could be exploited to learn domain-invariant stutter representations. These domain-invariant stutter-specific representations can, later on, be used to improve the performance of various ASIS systems.
\par 
One more issue with the ASIS systems is the generalization of trained models. Several techniques such as early stopping, regularization, and dropout have been used to improve generalization \cite{pironkov2016multi}. The main drawback of these techniques is that they are limited by the identification/recognition task. This problem can be solved by the multi-task learning strategy, (i.e., if the model is forced to learn some auxiliary tasks in parallel, in addition to its main task). Language and gender classifications are two auxiliary tasks, that can be learned together with the stutter identification task on the same input feature space to improve generalization. Recently, Sheikh \emph{et al}~\cite{stutternetmtl} used speaker recognition as a secondary task in learning robust stuttering representations by applying adversarial learning paradigm.  

\subsection{Multimodal Learning}
In stuttering identification, DL has been successfully applied to single modalities like text and audio. Inspired by the human brain, where the perceptions are carried out through the integration of information from several sensory inputs including vision, hearing, smell, etc., Ngiam \emph{et al}~\cite{ngiam2011multimodal} proposed a multi-modal (audio visual) learning and showed how to train deep models that learn effective shared representations across the modalities. The stuttering itself exhibits an audio-visual problem. Cues are present both in the visual (e.g. head nodding, lip tremors, quick eye blinks, and unusual lip shapes) as well as in the audio modality~\cite{guitar2013stuttering}. This multimodal learning paradigm could be helpful in learning robust stutter-specific hidden representations across the cross-modality platform, and could also help in building robust ASIS systems. Self-supervised learning can also be exploited to capture acoustic stutter-specific representations based on guided video frames. As proposed by~ Shukla \emph{et al}~\cite{shukla2020visually}, this framework could be helpful in learning stutter-specific features from audio signals guided by visual frames or vice-versa. Altinkaya and Smeulders~\cite{audiovisualdataset} recently presented the first audio-visual stuttered dataset which consists of 25 speakers (14 male, 11 female). They trained ResNet-based RNN (gated recurrent unit) on the audio-visual modality for the detection of block stuttering type. 

\subsection{Attention Approach}
The nature of stuttering is that it either usually happens on specific words or part of words, or part of sounds, syllables, or phrases and thus, is contained only in certain frames. The attention networks have been successfully applied in speech emotion detection~\cite{zhang2018attention}, speaker verification~\cite{zhang2016end}, speech recognition~\cite{chorowskiattention}. Attention networks~\cite{vaswani2017attention, chorowskiattention} which imitates cognitive attention can be exploited to force the neural networks to focus on the particular stutter-embedded frames that may improve the detection performance of ASIS systems.

\subsection{Multi-Stuttering Identification}
Most of the ASIS studies focus on utterances, which consist of only one type of stuttering.
However, the speech utterance can contain a mix of stuttering such as \textit{d-d-d-----dog dog is big}, which consists of syllable repetition, prolongation and word repetition types of disfluencies \cite{sawyer2010numbers}. There is a lack of studies in detecting multiple stuttering types if present in an utterance, and to the best of our knowledge, Ghonem \emph{et al}~\cite{ghonem2017classification} is the only study, that has been carried out to detect multiple stuttering types (\textit{repetition-prolongation}) in an utterance.

\section{Conclusion}
\label{Conclusion}
 Stuttering is a speech disorder during which the flow of speech is interrupted by involuntary blocks, prolongations, and repetitions. The conventional assessment of stuttering is to count manually
the occurrences of stuttering types and indicate them as a proportion to the total number of words in a speech
passage. The main drawback of this manual counting is that they are time-consuming and subjective which
makes it inconsistent and prone to error across different judges/STs. Approximately 70 million
people suffer from the stuttering problem worldwide which constitutes 1\% of the world’s population. Among
them, the stuttering is significant in males which is approximately four-fifth. 
\par 
Stuttering identification is a complex interdisciplinary problem that involves speech processing, signal processing, neuroscience, psychology, pathology, and machine learning. The recent advancements in the machine and DL have significantly transformed the speech domain. However, in SD, it has not been explored eminently. This work tries to fill the gap by trying to bring researchers together from interdisciplinary fields.  In the past two decades, a lot of research work has been performed on the automatic identification of stuttering. This paper gives an up-to-date comprehensive review of the various datasets, acoustic features, and ASIS classification methods, that have been used by various researchers for the identification and recognition of stuttering disfluencies. In this paper, We also discussed several challenges with possible solutions that need to be addressed for future work. These ASIS systems demand the training data among which the most common dataset, that has been used in the stuttering research is UCLASS \cite{howell2009university}.

\par 
Due to the challenges discussed in the \Cref{Challenges}, ASIS systems are not yet available for real-time stutter identification, unlike ASR, which are easily accessible on portable mobile devices. To achieve this goal, ASIS systems demand more powerful models so that the stuttering identification rate increases in cross-language and cross-speaker platforms with no labelled or very few annotated data.

\section*{Acknowledgements}
This  work  was  made  with  the  support  of  the  French  National  Research Agency, in the framework of the project ANR BENEPHIDIRE (18-CE36-0008-03). Experiments  presented  in  this  paper  were  carried  out  using  the  Grid’5000 testbed, supported by a scientific interest group hosted by Inria and including CNRS,  RENATER  and  several  universities  as  well  as  other  organizations(see  https://www.grid5000.fr) and  using the EXPLOR  centre, hosted by the University of Lorraine.

\bibliography{mybibfile}

\begin{thebibliography}{100}
\expandafter\ifx\csname url\endcsname\relax
  \def\url#1{\texttt{#1}}\fi
\expandafter\ifx\csname urlprefix\endcsname\relax\def\urlprefix{URL }\fi
\expandafter\ifx\csname href\endcsname\relax
  \def\href#1#2{#2} \def\path#1{#1}\fi

\bibitem{guitar2013stuttering}
B.~Guitar, Stuttering: An Integrated Approach to its Nature and Treatment,
  Lippincott Williams \& Wilkins, 2013.

\bibitem{duffy2013motor}
J.~R. Duffy, Motor Speech Disorders-E-Book: Substrates, Differential Diagnosis,
  and Management, Elsevier Health Sciences, 2013.

\bibitem{ratner2018fluency}
N.~B. Ratner, B.~MacWhinney, Fluency bank: A new resource for fluency research
  and practice, Journal of Fluency Disorders 56 (2018) 69.

\bibitem{ward2008stuttering}
D.~Ward, Stuttering and Cluttering: Frameworks for Understanding and Treatment,
  Psychology Press, 2008.

\bibitem{kehoe2006speech}
T.~D. Kehoe, W.~Contributors, Speech Language Pathology-Stuttering, Kiambo
  Ridge, 2006.

\bibitem{smith2017stuttering}
A.~Smith, C.~Weber, How stuttering develops: The multifactorial dynamic
  pathways theory, Journal of Speech, Language, and Hearing Research 60~(9)
  (2017) 2483--2505.

\bibitem{riva2008phenomenology}
P.~Riva-Posse, L.~Busto-Marolt, {\'A}.~Schteinschnaider, L.~Martinez-Echenique,
  {\'A}.~Cammarota, M.~Merello, Phenomenology of abnormal movements in
  stuttering, Parkinsonism \& Related Disorders 14~(5) (2008) 415--419.

\bibitem{starkweather1987fluency}
C.~W. Starkweather, Fluency and Stuttering., Prentice-Hall, Inc, 1987.

\bibitem{adams1974physiologic}
M.~R. Adams, A physiologic and aerodynamic interpretation of fluent and
  stuttered speech, Journal of Fluency Disorders 1~(1) (1974) 35--47.

\bibitem{roberts2009disfluencies}
P.~M. Roberts, A.~Meltzer, J.~Wilding, Disfluencies in non-stuttering adults
  across sample lengths and topics, Journal of Communication Disorders 42~(6)
  (2009) 414--427.

\bibitem{world1977manual}
W.~H. Organization, et~al., Manual of the international statistical
  classification of diseases, injuries, and causes of death: based on the
  recommendations of the ninth revision conference, 1975, and adopted by the
  Twenty-ninth World Health Assembly, World Health Organization, 1977.

\bibitem{etchell2018systematic}
A.~C. Etchell, O.~Civier, K.~J. Ballard, P.~F. Sowman, A systematic literature
  review of neuroimaging research on developmental stuttering between 1995 and
  2016, Journal of Fluency Disorders 55 (2018) 6--45.

\bibitem{drayna2011genetic}
D.~Drayna, C.~Kang, Genetic approaches to understanding the causes of
  stuttering, Journal of Neurodevelopmental Disorders 3~(4) (2011) 374--380.

\bibitem{vanhoutte2016will}
S.~Vanhoutte, M.~Cosyns, P.~van Mierlo, K.~Batens, P.~Corthals, M.~De~Letter,
  J.~Van~Borsel, P.~Santens, When will a stuttering moment occur? the
  determining role of speech motor preparation, Neuropsychologia 86 (2016)
  93--102.

\bibitem{neef2015speech}
N.~E. Neef, T.~L. Hoang, A.~Neef, W.~Paulus, M.~Sommer, Speech dynamics are
  coded in the left motor cortex in fluent speakers but not in adults who
  stutter, Brain 138~(3) (2015) 712--725.

\bibitem{belyk2015stuttering}
M.~Belyk, S.~J. Kraft, S.~Brown, Stuttering as a trait or state--an ale
  meta-analysis of neuroimaging studies, European Journal of Neuroscience
  41~(2) (2015) 275--284.

\bibitem{riaz2005genomewide}
N.~Riaz, S.~Steinberg, J.~Ahmad, A.~Pluzhnikov, S.~Riazuddin, N.~J. Cox,
  D.~Drayna, Genomewide significant linkage to stuttering on chromosome 12, The
  American Journal of Human Genetics 76~(4) (2005) 647--651.

\bibitem{stuttering}
NIDCD, \href{: https://www.nidcd.nih.gov/health/stuttering/}{Stuttering}.
\newline\urlprefix\url{: https://www.nidcd.nih.gov/health/stuttering/}

\bibitem{yairi2013epidemiology}
E.~Yairi, N.~Ambrose, Epidemiology of stuttering: 21st century advances,
  Journal of Fluency Disorders 38~(2) (2013) 66--87.

\bibitem{iverach2016prevalence}
L.~Iverach, M.~Jones, L.~F. McLellan, H.~J. Lyneham, R.~G. Menzies, M.~Onslow,
  R.~M. Rapee, Prevalence of anxiety disorders among children who stutter,
  Journal of Fluency Disorders 49 (2016) 13--28.

\bibitem{national2009experience}
N.~S.~A. NSA, The experience of people who stutter: A survey by the national
  stuttering association, New York, NY: Author.

\bibitem{klein2004impact}
J.~F. Klein, S.~B. Hood, The impact of stuttering on employment opportunities
  and job performance, Journal of Fluency Disorders 29~(4) (2004) 255--273.

\bibitem{blood2016long}
G.~W. Blood, I.~M. Blood, Long-term consequences of childhood bullying in
  adults who stutter: Social anxiety, fear of negative evaluation, self-esteem,
  and satisfaction with life, Journal of Fluency Disorders 50 (2016) 72--84.

\bibitem{antipova2008effects}
E.~A. Antipova, S.~C. Purdy, M.~Blakeley, S.~Williams, Effects of altered
  auditory feedback (aaf) on stuttering frequency during monologue speech
  production, Journal of Fluency Disorders 33~(4) (2008) 274--290.

\bibitem{noth2000automatic}
E.~N{\"o}th, H.~Niemann, T.~Haderlein, M.~Decher, U.~Eysholdt, F.~Rosanowski,
  T.~Wittenberg, Automatic stuttering recognition using hidden {Markov} models,
  in: Proc. Sixth International Conference on Spoken Language Processing, 2000.

\bibitem{saltuklaroglu2005effective}
T.~Saltuklaroglu, J.~Kalinowski, How effective is therapy for childhood
  stuttering? dissecting and reinterpreting the evidence in light of
  spontaneous recovery rates, International Journal of Language \&
  Communication Disorders 40~(3) (2005) 359--374.

\bibitem{roberts2011using}
M.~Y. Roberts, Using emprical benchmarks to assess the effects of a
  parentimplemented language intervention for children with language
  impairments, Vanderbilt University, 2011.

\bibitem{usatodaytech2020}
U.~T. TECH, For people who stutter, the convenience of voice assistant
  technology remains out of reach,
  \url{https://eu.usatoday.com/story/tech/2020/01/06/voice-assistants-remain-out-reach-people-who-stutter/2749115001/},
  accessed: 2020-12-24.

\bibitem{kourkounakis2020detecting}
T.~Kourkounakis, A.~Hajavi, A.~Etemad, Detecting multiple speech disfluencies
  using a deep residual network with bidirectional long short-term memory, in:
  Proc. ICASSP 2020-2020 IEEE International Conference on Acoustics, Speech and
  Signal Processing (ICASSP), IEEE, 2020, pp. 6089--6093.

\bibitem{sheikh:hal-03227223}
S.~A. Sheikh, M.~Sahidullah, F.~Hirsch, S.~Ouni,
  \href{https://hal.inria.fr/hal-03227223}{{StutterNet: Stuttering Detection
  Using Time Delay Neural Network}}, in: {Proc. EUSIPCO 2021 -- 29th European
  Signal Processing Conference}, Dublin, Ireland, 2021.
\newline\urlprefix\url{https://hal.inria.fr/hal-03227223}

\bibitem{stutternetmtl}
S.~A. Sheikh, et~al., Robust stuttering detection via multi-task and
  adversarial learning, in: Proc. 30th EUSIPCO, 2022.

\bibitem{bayerw2v2}
S.~P. Bayerl, D.~Wagner, E.~N{\"o}th, K.~Riedhammer, Detecting dysfluencies in
  stuttering therapy using wav2vec 2.0, arXiv preprint arXiv:2204.03417.

\bibitem{kourkounakis2020fluentnet}
T.~Kourkounakis, A.~Hajavi, A.~Etemad, Fluentnet: End-to-end detection of
  speech disfluency with deep learning, arXiv preprint arXiv:2009.11394.

\bibitem{sep28k}
C.~Lea, V.~Mitra, A.~Joshi, S.~Kajarekar, J.~P. Bigham, Sep-28k: A dataset for
  stuttering event detection from podcasts with people who stutter, in: Proc.
  ICASSP 2021 - 2021 IEEE International Conference on Acoustics, Speech and
  Signal Processing (ICASSP), 2021, pp. 6798--6802.
\newblock \href {http://dx.doi.org/10.1109/ICASSP39728.2021.9413520}
  {\path{doi:10.1109/ICASSP39728.2021.9413520}}.

\bibitem{sheikhw2v2}
S.~A. Sheikh, M.~Sahidullah, F.~Hirsch, S.~Ouni, Introducing {ECAPA-TDNN} and
  {Wav2Vec2.0} embeddings to stuttering detection, arXiv preprint
  arXiv:2204.01564.

\bibitem{nassif2019speech}
A.~B. Nassif, I.~Shahin, I.~Attili, M.~Azzeh, K.~Shaalan, Speech recognition
  using deep neural networks: A systematic review, IEEE access 7 (2019)
  19143--19165.

\bibitem{akccay2020speech}
M.~B. Ak{\c{c}}ay, K.~O{\u{g}}uz, Speech emotion recognition: Emotional models,
  databases, features, preprocessing methods, supporting modalities, and
  classifiers, Speech Communication 116 (2020) 56--76.

\bibitem{ning2019review}
Y.~Ning, S.~He, Z.~Wu, C.~Xing, L.-J. Zhang, A review of deep learning based
  speech synthesis, Applied Sciences 9~(19) (2019) 4050.

\bibitem{yildirim2009automatic}
S.~Yildirim, S.~Narayanan, Automatic detection of disfluency boundaries in
  spontaneous speech of children using audio--visual information, IEEE
  Transactions on Audio, Speech, and Language Processing 17~(1) (2009) 2--12.

\bibitem{geetha2000classification}
Y.~Geetha, K.~Pratibha, R.~Ashok, S.~K. Ravindra, Classification of childhood
  disfluencies using neural networks, Journal of Fluency Disorders 25~(2)
  (2000) 99--117.

\bibitem{villegas2019novel}
B.~Villegas, K.~M. Flores, K.~J. Acu{\~n}a, K.~Pacheco-Barrios, D.~Elias, A
  novel stuttering disfluency classification system based on respiratory
  biosignals, in: Proc. 2019 41st Annual International Conference of the IEEE
  Engineering in Medicine and Biology Society (EMBC), IEEE, 2019, pp.
  4660--4663.

\bibitem{fmri}
R.~Hosseini, B.~Walsh, F.~Tian, S.~Wang, An f{NIRS}-based feature learning and
  classification framework to distinguish hemodynamic patterns in children who
  stutter, IEEE Transactions on Neural Systems and Rehabilitation Engineering
  26~(6) (2018) 1254--1263.
\newblock \href {http://dx.doi.org/10.1109/TNSRE.2018.2829083}
  {\path{doi:10.1109/TNSRE.2018.2829083}}.

\bibitem{chang2014research}
S.-E. Chang, Research updates in neuroimaging studies of children who stutter,
  in: Seminars in Speech and Language, Vol.~35, NIH Public Access, 2014, p.~67.

\bibitem{ingham1996functional}
R.~J. Ingham, P.~T. Fox, J.~C. Ingham, F.~Zamarripa, C.~Martin, P.~Jerabek,
  J.~Cotton, Functional-lesion investigation of developmental stuttering with
  positron emission tomography, Journal of Speech, Language, and Hearing
  Research 39~(6) (1996) 1208--1227.

\bibitem{foundas2001anomalous}
A.~Foundas, A.~Lane, D.~Corey, M.~Hurley, K.~Heilman, Anomalous anatomy in
  adults with persistent developmental stuttering: A volumetric mri study of
  cortical speech and language areas, in: Neurology, Vol.~56, 2001, pp.
  A157--A158.

\bibitem{conture1977laryngeal}
E.~G. Conture, G.~N. McCall, D.~W. Brewer, Laryngeal behavior during
  stuttering, Journal of Speech and Hearing Research 20~(4) (1977) 661--668.

\bibitem{conture1985laryngeal}
E.~G. Conture, H.~D. Schwartz, D.~W. Brewer, Laryngeal behavior during
  stuttering: A further study, Journal of Speech, Language, and Hearing
  Research 28~(2) (1985) 233--240.

\bibitem{wingate1969stuttering}
M.~E. Wingate, Stuttering as phonetic transition defect, Journal of Speech and
  Hearing Disorders 34~(1) (1969) 107--108.

\bibitem{didirkova2020two}
I.~Didirkov{\'a}, F.~Hirsch, A two-case study of coarticulation in stuttered
  speech. an articulatory approach, Clinical Linguistics \& Phonetics 34~(6)
  (2020) 517--535.

\bibitem{didirkova2020articulatory}
I.~Didirkova, S.~Le~Maguer, F.~Hirsch, An articulatory study of differences and
  similarities between stuttered disfluencies and non-pathological
  disfluencies, Clinical Linguistics \& Phonetics (2020) 1--21.

\bibitem{jayaram1983phonetic}
M.~Jayaram, Phonetic influences on stuttering in monolingual and bilingual
  stutterers, Journal of Communication Disorders 16~(4) (1983) 287--297.

\bibitem{blomgren2012speech}
M.~Blomgren, M.~Alqhazo, E.~Metzger, Do speech sound characteristics really
  influence stuttering frequency, in: Proc. of the 7th World Congress of
  Fluency Disorders, CD-ROM, 2012.

\bibitem{didirkova2016parole}
I.~Didirkova, Parole, langues et disfluences: une {\'e}tude linguistique et
  phon{\'e}tique du b{\'e}gaiement, Ph.D. thesis, Universit{\'e} Paul
  Val{\'e}ry-Montpellier III (2016).

\bibitem{zebrowski1985acoustic}
P.~M. Zebrowski, E.~G. Conture, E.~A. Cudahy, Acoustic analysis of young
  stutterers' fluency: Preliminary observations, Journal of Fluency Disorders
  10~(3) (1985) 173--192.

\bibitem{dehqan2016formant}
A.~Dehqan, F.~Yadegari, M.~Blomgren, R.~C. Scherer, Formant transitions in the
  fluent speech of {F}arsi-speaking people who stutter, Journal of Fluency
  Disorders 48 (2016) 1--15.

\bibitem{yaruss1993f2}
J.~S. Yaruss, E.~G. Conture, F2 transitions during sound/syllable repetitions
  of children who stutter and predictions of stuttering chronicity, Journal of
  Speech, Language, and Hearing Research 36~(5) (1993) 883--896.

\bibitem{robb1998formant}
M.~Robb, M.~Blomgren, Y.~Chen, Formant frequency fluctuation in stuttering and
  nonstuttering adults, Journal of Fluency Disorders 23~(1) (1998) 73--84.

\bibitem{chang2002coarticulation}
S.-E. Chang, R.~N. Ohde, E.~G. Conture, Coarticulation and formant transition
  rate in young children who stutter, Journal of Speech, Language, and Hearing
  Research.

\bibitem{subramanian2003second}
A.~Subramanian, E.~Yairi, O.~Amir, Second formant transitions in fluent speech
  of persistent and recovered preschool children who stutter, Journal of
  Communication Disorders 36~(1) (2003) 59--75.

\bibitem{blomgren1998note}
M.~Blomgren, M.~Robb, Y.~Chen, A note on vowel centralization in stuttering and
  nonstuttering individuals, Journal of Speech, Language, and Hearing Research
  41~(5) (1998) 1042--1051.

\bibitem{hirsch:halshs-00716583}
F.~Hirsch, F.~Bouarourou, B.~Vaxelaire, M.-C. Monfrais-Pfauwadel, M.~Bechet,
  J.~Sturm, R.~Sock,
  \href{https://halshs.archives-ouvertes.fr/halshs-00716583}{Formant structures
  of vowels produced by stutterers in normal and fast speech rates}, in: {8th
  International Seminar On Speech Production}, France, 2008, p.~NC.
\newline\urlprefix\url{https://halshs.archives-ouvertes.fr/halshs-00716583}

\bibitem{healey1986acoustic}
E.~C. Healey, P.~R. Ramig, Acoustic measures of stutterers' and nonstutterers'
  fluency in two speech contexts, Journal of Speech, Language, and Hearing
  Research 29~(3) (1986) 325--331.

\bibitem{hillman1977voice}
R.~E. Hillman, H.~R. Gilbert, Voice onset time for voiceless stop consonants in
  the fluent reading of stutterers and nonstutterers, The Journal of the
  Acoustical Society of America 61~(2) (1977) 610--611.

\bibitem{adams1987voice}
M.~R. Adams, Voice onsets and segment durations of normal speakers and
  beginning stutterers, Journal of Fluency Disorders 12~(2) (1987) 133--139.

\bibitem{watson1982comparison}
B.~C. Watson, P.~J. Alfonso, A comparison of lrt and vot values between
  stutterers and nonstutterers, Journal of Fluency Disorders 7~(2) (1982)
  219--241.

\bibitem{jancke1994variability}
L.~J{\"a}ncke, Variability and duration of voice onset time and phonation in
  stuttering and nonstuttering adults, Journal of Fluency Disorders 19~(1)
  (1994) 21--37.

\bibitem{de1991voice}
L.~F. De~Nil, G.~Brutten, Voice onset times of stuttering and nonstuttering
  children: The influence of externally and linguistically imposed time
  pressure, Journal of Fluency Disorders 16~(2-3) (1991) 143--158.

\bibitem{celeste2015impact}
L.~C. Celeste, V.~de~Oliveira Martins-Reis, The impact of a dysfluency
  environment on the temporal organization of consonants in stuttering,
  Audiology-Communication Research 20~(1) (2015) 10--17.

\bibitem{brosch2002prognostic}
S.~Brosch, A.~H{\"a}ge, H.~S. Johannsen, Prognostic indicators for stuttering:
  The value of computer-based speech analysis, Brain and Language 82~(1) (2002)
  75--86.

\bibitem{borden1985onset}
G.~J. Borden, T.~Baer, M.~K. Kenney, Onset of voicing in stuttered and fluent
  utterances, Journal of Speech, Language, and Hearing Research 28~(3) (1985)
  363--372.

\bibitem{fosnot1999prosodic}
S.~M. Fosnot, S.~Jun, Prosodic characteristics in children with stuttering or
  autism during reading and imitation, in: Proc. of the 14th International
  Congress of Phonetic Sciences, 1999, pp. 1925--1928.

\bibitem{ramig1981vocal}
P.~R. Ramig, M.~R. Adams, Vocal changes in stutterers and nonstutterers during
  high-and low-pitched speech, Journal of Fluency Disorders 6~(1) (1981)
  15--33.

\bibitem{howell2009university}
P.~Howell, S.~Davis, J.~Bartrip, The university college {L}ondon archive of
  stuttered speech ({UCLASS}), Journal of Speech, Language, and Hearing
  Research.

\bibitem{rudzicz2012torgo}
F.~Rudzicz, A.~K. Namasivayam, T.~Wolff, The torgo database of acoustic and
  articulatory speech from speakers with dysarthria, Journal of Language
  Resources and Evaluation 46~(4) (2012) 523--541.

\bibitem{bayerl2022ksof}
S.~P. Bayerl, A.~W. von Gudenberg, F.~H{\"o}nig, E.~N{\"o}th, K.~Riedhammer,
  {KSoF: The Kassel state of fluency dataset--A therapy centered dataset of
  stuttering}, arXiv preprint arXiv:2203.05383.

\bibitem{Zayats2016}
V.~Zayats, M.~Ostendorf, H.~Hajishirzi,
  \href{http://dx.doi.org/10.21437/Interspeech.2016-1247}{Disfluency detection
  using a bidirectional {LSTM}}, in: Proc. Interspeech 2016, 2016, pp.
  2523--2527.
\newblock \href {http://dx.doi.org/10.21437/Interspeech.2016-1247}
  {\path{doi:10.21437/Interspeech.2016-1247}}.
\newline\urlprefix\url{http://dx.doi.org/10.21437/Interspeech.2016-1247}

\bibitem{chen}
Q.~Chen, M.~Chen, B.~Li, W.~Wang, Controllable time-delay transformer for
  real-time punctuation prediction and disfluency detection, in: Proc. ICASSP
  2020 - 2020 IEEE International Conference on Acoustics, Speech and Signal
  Processing (ICASSP), 2020, pp. 8069--8073.
\newblock \href {http://dx.doi.org/10.1109/ICASSP40776.2020.9053159}
  {\path{doi:10.1109/ICASSP40776.2020.9053159}}.

\bibitem{Alharbi2018}
S.~Alharbi, M.~Hasan, A.~{J H Simons}, S.~Brumfitt, P.~Green,
  \href{http://dx.doi.org/10.21437/Interspeech.2018-2155}{A lightly supervised
  approach to detect stuttering in children's speech}, in: Proc. Interspeech
  2018, 2018, pp. 3433--3437.
\newblock \href {http://dx.doi.org/10.21437/Interspeech.2018-2155}
  {\path{doi:10.21437/Interspeech.2018-2155}}.
\newline\urlprefix\url{http://dx.doi.org/10.21437/Interspeech.2018-2155}

\bibitem{ALHARBI2020101052}
S.~Alharbi, M.~Hasan, A.~J.~H. Simons, S.~Brumfitt, P.~Green,
  \href{https://www.sciencedirect.com/science/article/pii/S0885230819302967}{Sequence
  labeling to detect stuttering events in read speech}, Computer Speech \&
  Language 62 (2020) 101052.
\newblock \href {http://dx.doi.org/https://doi.org/10.1016/j.csl.2019.101052}
  {\path{doi:https://doi.org/10.1016/j.csl.2019.101052}}.
\newline\urlprefix\url{https://www.sciencedirect.com/science/article/pii/S0885230819302967}

\bibitem{huang2001spoken}
X.~Huang, A.~Acero, H.-W. Hon, R.~Reddy, Spoken Language Processing: A Guide to
  Theory, Algorithm, and System Development, Prentice hall PTR, 2001.

\bibitem{howell1995automatic}
P.~Howell, S.~Sackin, Automatic recognition of repetitions and prolongations in
  stuttered speech, in: Proc. of the first World Congress on Fluency Disorders,
  Vol.~2, University Press Nijmegen Nijmegen, The Netherlands, 1995, pp.
  372--374.

\bibitem{howell1997development1}
P.~Howell, S.~Sackin, K.~Glenn, Development of a two-stage procedure for the
  automatic recognition of dysfluencies in the speech of children who stutter:
  I. psychometric procedures appropriate for selection of training material for
  lexical dysfluency classifiers, Journal of Speech, Language, and Hearing
  Research 40~(5) (1997) 1073--1084.

\bibitem{howell1997development2}
P.~Howell, S.~Sackin, K.~Glenn, Development of a two-stage procedure for the
  automatic recognition of dysfluencies in the speech of children who stutter:
  Ii. ann recognition of repetitions and prolongations with supplied word
  segment markers, Journal of Speech, Language, and Hearing Research 40~(5)
  (1997) 1085--1096.

\bibitem{khara2018comparative}
S.~Khara, S.~Singh, D.~Vir, A comparative study of the techniques for feature
  extraction and classification in stuttering, in: Proc. 2018 Second
  International Conference on Inventive Communication and Computational
  Technologies (ICICCT), IEEE, 2018, pp. 887--893.

\bibitem{czyzewski2003intelligent}
A.~Czyzewski, A.~Kaczmarek, B.~Kostek, Intelligent processing of stuttered
  speech, Journal of Intelligent Information Systems 21~(2) (2003) 143--171.

\bibitem{szczurowska2014application}
I.~Szczurowska, W.~Kuniszyk-J{\'o}{\'z}kowiak, E.~Smo{\l}ka, The application of
  kohonen and multilayer perceptron networks in the speech nonfluency analysis,
  Archives of Acoustics 31~(4 (S)) (2014) 205--210.

\bibitem{swietlicka2009artificial}
I.~{\'S}wietlicka, W.~Kuniszyk-J{\'o}{\'z}kowiak, E.~Smo{\l}ka, Artificial
  neural networks in the disabled speech analysis, in: Computer Recognition
  Systems 3, Springer, 2009, pp. 347--354.

\bibitem{chee2009overview}
L.~S. Chee, O.~C. Ai, S.~Yaacob, Overview of automatic stuttering recognition
  system, in: Proc. International Conference on Man-Machine Systems, no.
  October, Batu Ferringhi, Penang Malaysia, 2009, pp. 1--6.

\bibitem{hariharan2012classification}
M.~Hariharan, L.~S. Chee, O.~C. Ai, S.~Yaacob, Classification of speech
  dysfluencies using {LPC} based parameterization techniques, Journal of
  Medical Systems 36~(3) (2012) 1821--1830.

\bibitem{esmaili2017automatic}
I.~Esmaili, N.~J. Dabanloo, M.~Vali, An automatic prolongation detection
  approach in continuous speech with robustness against speaking rate
  variations, Journal of Medical Signals and Sensors 7 (2017) 1.

\bibitem{lopez2018analysis}
K.~L{\'o}pez-de Ipi{\~n}a, U.~Martinez-de Lizarduy, P.~Calvo, B.~Beitia,
  J.~Garc{\'\i}a-Melero, E.~Fern{\'a}ndez, M.~Ecay-Torres, M.~Faundez-Zanuy,
  P.~Sanz, On the analysis of speech and disfluencies for automatic detection
  of mild cognitive impairment, Neural Computing and Applications (2018) 1--9.

\bibitem{mahesha2017lp}
P.~Mahesha, D.~Vinod, {LP-Hillbert transform based MFCC for effective
  discrimination of stuttering dysfluencies}, in: Proc. 2017 International
  Conference on Wireless Communications, Signal Processing and Networking
  (WiSPNET), IEEE, 2017, pp. 2561--2565.

\bibitem{ghonem2017classification}
S.~A. Ghonem, S.~Abdou, M.~A. Esmael, N.~Ghamry, Classification of stuttering
  events using i-vector, The Egyptian Journal of Language Engineering 4~(1)
  (2017) 11--19.

\bibitem{hariharan2012speech}
M.~Hariharan, V.~Vijean, C.~Fook, S.~Yaacob, Speech stuttering assessment using
  sample entropy and least square support vector machine, in: Proc. 2012 IEEE
  8th International Colloquium on Signal Processing and its Applications, IEEE,
  2012, pp. 240--245.

\bibitem{mahesha2013classification}
P.~Mahesha, D.~Vinod, Classification of speech dysfluencies using speech
  parameterization techniques and multiclass svm, in: Proc. International
  Conference on Heterogeneous Networking for Quality, Reliability, Security and
  Robustness, Springer, 2013, pp. 298--308.

\bibitem{fook2013comparison}
C.~Y. Fook, H.~Muthusamy, L.~S. Chee, S.~B. Yaacob, A.~H.~B. Adom, Comparison
  of speech parameterization techniques for the classification of speech
  disfluencies, Turkish Journal of Electrical Engineering \& Computer Sciences
  21~(Sup. 1) (2013) 1983--1994.

\bibitem{arjun2020automatic}
K.~Arjun, S.~Karthik, D.~Kamalnath, P.~Chanda, S.~Tripathi, Automatic
  correction of stutter in disfluent speech, Procedia Computer Science 171
  (2020) 1363--1370.

\bibitem{ai2012classification}
O.~C. Ai, M.~Hariharan, S.~Yaacob, L.~S. Chee, {Classification of speech
  dysfluencies with MFCC and LPCC features}, Expert Systems with Applications
  39~(2) (2012) 2157--2165.

\bibitem{suszynski2015prolongation}
W.~Suszy{\'n}ski, W.~Kuniszyk-J{\'o}{\'z}kowiak, E.~Smo{\l}ka,
  M.~Dzie{\'n}kowski, Prolongation detection with application of fuzzy logic,
  Annales Universitatis Mariae Curie-Sklodowska, sectio AI--Informatica 1~(1)
  (2015) 1--8.

\bibitem{wisniewski2007automatic}
M.~Wi{\'s}niewski, W.~Kuniszyk-J{\'o}{\'z}kowiak, E.~Smo{\l}ka,
  W.~Suszy{\'n}ski, Automatic detection of disorders in a continuous speech
  with the hidden markov models approach, in: Computer Recognition Systems 2,
  Springer, 2007, pp. 445--453.

\bibitem{tan2007application}
T.-S. Tan, A.~Ariff, C.-M. Ting, S.-H. Salleh, et~al., Application of malay
  speech technology in malay speech therapy assistance tools, in: Proc. 2007
  International Conference on Intelligent and Advanced Systems, IEEE, 2007, pp.
  330--334.

\bibitem{ravikumar2008automatic}
K.~Ravikumar, B.~Reddy, R.~Rajagopal, H.~Nagaraj, Automatic detection of
  syllable repetition in read speech for objective assessment of stuttered
  disfluencies, Proc. of World Academy of Science, Engineering and Technology
  36 (2008) 270--273.

\bibitem{chee2009mfcc}
L.~S. Chee, O.~C. Ai, M.~Hariharan, S.~Yaacob, {MFCC based recognition of
  repetitions and prolongations in stuttered speech using k-NN and LDA}, in:
  Proc. 2009 IEEE Student Conference on Research and Development (SCOReD),
  IEEE, 2009, pp. 146--149.

\bibitem{chee2009automatic}
L.~S. Chee, O.~C. Ai, M.~Hariharan, S.~Yaacob, Automatic detection of
  prolongations and repetitions using {LPCC}, in: Proc. 2009 international
  conference for technical postgraduates (TECHPOS), IEEE, 2009, pp. 1--4.

\bibitem{ravikumar2009approach}
K.~Ravikumar, R.~Rajagopal, H.~Nagaraj, {An approach for objective assessment
  of stuttered speech using MFCC}, in: Proc. The International Congress for
  Global Science and Technology, 2009, p.~19.

\bibitem{palfy2011recognition}
J.~P{\'a}lfy, J.~Posp{\'\i}chal, Recognition of repetitions using support
  vector machines, in: Signal Processing Algorithms, Architectures,
  Arrangements, and Applications SPA 2011, IEEE, 2011, pp. 1--6.

\bibitem{swietlicka2013hierarchical}
I.~{\'S}wietlicka, W.~Kuniszyk-J{\'o}{\'z}kowiak, E.~Smo{\l}ka, Hierarchical
  ann system for stuttering identification, Computer Speech \& Language 27~(1)
  (2013) 228--242.

\bibitem{oue2015automatic}
S.~Oue, R.~Marxer, F.~Rudzicz, Automatic dysfluency detection in dysarthric
  speech using deep belief networks, in: Proc. of SLPAT 2015: 6th Workshop on
  Speech and Language Processing for Assistive Technologies, 2015, pp. 60--64.

\bibitem{santoso2019classification}
J.~Santoso, T.~Yamada, S.~Makino, Classification of causes of speech
  recognition errors using attention-based bidirectional long short-term memory
  and modulation spectrum, in: Proc. 2019 Asia-Pacific Signal and Information
  Processing Association Annual Summit and Conference (APSIPA ASC), IEEE, 2019,
  pp. 302--306.

\bibitem{santoso2019categorizing}
J.~Santoso, T.~Yamada, S.~Makino, Categorizing error causes related to
  utterance characteristics in speech recognition, Proc. NCSP 19 (2019)
  514--517.

\bibitem{goodfellow2016deep}
I.~Goodfellow, Y.~Bengio, A.~Courville, Y.~Bengio, Deep learning, Vol.~1, MIT
  press Cambridge, 2016.

\bibitem{murphy2012machine}
K.~P. Murphy, Machine Learning: a Probabilistic Perspective, MIT press, 2012.

\bibitem{lstmorig}
S.~Hochreiter, J.~Schmidhuber, Long short-term memory, Neural Comput. 9~(8)
  (1997) 1735–1780.
\newblock \href {http://dx.doi.org/10.1162/neco.1997.9.8.1735}
  {\path{doi:10.1162/neco.1997.9.8.1735}}.

\bibitem{melanie}
M.~Jouaiti, K.~Dautenhahn, Dysfluency classification in stuttered speech using
  deep learning for real-time applications, in: Proc. ICASSP 2022.

\bibitem{yin2015noisy}
S.~Yin, C.~Liu, Z.~Zhang, Y.~Lin, D.~Wang, J.~Tejedor, T.~F. Zheng, Y.~Li,
  Noisy training for deep neural networks in speech recognition, EURASIP
  Journal on Audio, Speech, and Music Processing 2015~(1) (2015) 1--14.

\bibitem{latif2020deep}
S.~Latif, R.~Rana, S.~Khalifa, R.~Jurdak, J.~Qadir, B.~W. Schuller, Deep
  representation learning in speech processing: Challenges, recent advances,
  and future trends, arXiv preprint arXiv:2001.00378.

\bibitem{qian2019data}
Y.~Qian, H.~Hu, T.~Tan, Data augmentation using generative adversarial networks
  for robust speech recognition, Speech Communication 114 (2019) 1--9.

\bibitem{oord2018representation}
A.~v.~d. Oord, Y.~Li, O.~Vinyals, Representation learning with contrastive
  predictive coding, arXiv preprint arXiv:1807.03748.

\bibitem{zhu2009introduction}
X.~Zhu, A.~B. Goldberg, Introduction to semi-supervised learning, Synthesis
  lectures on artificial intelligence and machine learning 3~(1) (2009) 1--130.

\bibitem{shukla2020visually}
A.~Shukla, K.~Vougioukas, P.~Ma, S.~Petridis, M.~Pantic, Visually guided self
  supervised learning of speech representations, in: Proc. ICASSP 2020-2020
  IEEE International Conference on Acoustics, Speech and Signal Processing
  (ICASSP), IEEE, 2020, pp. 6299--6303.

\bibitem{chawla2002smote}
N.~V. Chawla, K.~W. Bowyer, L.~O. Hall, W.~P. Kegelmeyer, Smote: synthetic
  minority over-sampling technique, Journal of Artificial Intelligence Research
  16 (2002) 321--357.

\bibitem{cuiclassbalance}
Y.~Cui, M.~Jia, T.-Y. Lin, Y.~Song, S.~Belongie, Class-balanced loss based on
  effective number of samples, in: Proc. 2019 IEEE/CVF Conference on Computer
  Vision and Pattern Recognition (CVPR), 2019, pp. 9260--9269.
\newblock \href {http://dx.doi.org/10.1109/CVPR.2019.00949}
  {\path{doi:10.1109/CVPR.2019.00949}}.

\bibitem{wang2018iterative}
N.~Wang, X.~Zhao, Y.~Jiang, Y.~Gao, K.~BNRist, Iterative metric learning for
  imbalance data classification., in: Proc. IJCAI, 2018, pp. 2805--2811.

\bibitem{yang2020rethinking}
Y.~Yang, Z.~Xu, Rethinking the value of labels for improving class-imbalanced
  learning, in: H.~Larochelle, M.~Ranzato, R.~Hadsell, M.~F. Balcan, H.~Lin
  (Eds.), Advances in Neural Information Processing Systems, Vol.~33, Curran
  Associates, Inc., 2020, pp. 19290--19301.

\bibitem{Xuechen9401593}
X.~Liu, M.~Sahidullah, T.~Kinnunen, Learnable {MFCCs} for speaker verification,
  in: Proc. 2021 IEEE International Symposium on Circuits and Systems (ISCAS),
  2021, pp. 1--5.
\newblock \href {http://dx.doi.org/10.1109/ISCAS51556.2021.9401593}
  {\path{doi:10.1109/ISCAS51556.2021.9401593}}.

\bibitem{sailor2016unsupervised}
H.~B. Sailor, H.~A. Patil, Unsupervised deep auditory model using stack of
  convolutional {RBMs} for speech recognition., in: Proc. INTERSPEECH, 2016,
  pp. 3379--3383.

\bibitem{millet2019learning}
J.~Millet, N.~Zeghidour, Learning to detect dysarthria from raw speech, in:
  Proc. ICASSP 2019-2019 IEEE International Conference on Acoustics, Speech and
  Signal Processing (ICASSP), IEEE, 2019, pp. 5831--5835.

\bibitem{daemotion}
M.~Abdelwahab, C.~Busso, Supervised domain adaptation for emotion recognition
  from speech, in: Proc. 2015 IEEE International Conference on Acoustics,
  Speech and Signal Processing (ICASSP), 2015, pp. 5058--5062.
\newblock \href {http://dx.doi.org/10.1109/ICASSP.2015.7178934}
  {\path{doi:10.1109/ICASSP.2015.7178934}}.

\bibitem{Alam2018}
M.~J. Alam, G.~Bhattacharya, P.~Kenny,
  \href{http://dx.doi.org/10.21437/Odyssey.2018-25}{Speaker verification in
  mismatched conditions with frustratingly easy domain adaptation}, in: Proc.
  Odyssey 2018 The Speaker and Language Recognition Workshop, 2018, pp.
  176--180.
\newblock \href {http://dx.doi.org/10.21437/Odyssey.2018-25}
  {\path{doi:10.21437/Odyssey.2018-25}}.
\newline\urlprefix\url{http://dx.doi.org/10.21437/Odyssey.2018-25}

\bibitem{samarakonASR}
L.~Samarakoon, B.~Mak, A.~Y. Lam, Domain adaptation of end-to-end speech
  recognition in low-resource settings, in: Proc. 2018 IEEE Spoken Language
  Technology Workshop (SLT), 2018, pp. 382--388.
\newblock \href {http://dx.doi.org/10.1109/SLT.2018.8639506}
  {\path{doi:10.1109/SLT.2018.8639506}}.

\bibitem{ganin2015unsupervised}
Y.~Ganin, V.~Lempitsky, Proc. unsupervised domain adaptation by
  backpropagation, in: International Conference on Machine Learning, PMLR,
  2015, pp. 1180--1189.

\bibitem{wang}
Q.~Wang, W.~Rao, S.~Sun, L.~Xie, E.~S. Chng, H.~Li, Unsupervised domain
  adaptation via domain adversarial training for speaker recognition, in: Proc.
  2018 IEEE International Conference on Acoustics, Speech and Signal Processing
  (ICASSP), 2018, pp. 4889--4893.
\newblock \href {http://dx.doi.org/10.1109/ICASSP.2018.8461423}
  {\path{doi:10.1109/ICASSP.2018.8461423}}.

\bibitem{pironkov2016multi}
G.~Pironkov, S.~Dupont, T.~Dutoit, Multi-task learning for speech recognition:
  an overview., in: Proc. European Symposium on Artificial Neural Networks,
  Computational Intelligence and Machine Learning (ESANN), 2016.

\bibitem{ngiam2011multimodal}
J.~Ngiam, A.~Khosla, M.~Kim, J.~Nam, H.~Lee, A.~Y. Ng, Proc. multimodal deep
  learning, in: ICML, 2011.

\bibitem{audiovisualdataset}
M.~Altinkaya, A.~W. Smeulders, \href{https://doi.org/10.1145/3423325.3423733}{A
  dynamic, self supervised, large scale audiovisual dataset for stuttered
  speech}, in: Proc. of the 1st International Workshop on Multimodal
  Conversational AI, MuCAI20, Association for Computing Machinery, New York,
  NY, USA, 2020, p. 9–13.
\newline\urlprefix\url{https://doi.org/10.1145/3423325.3423733}

\bibitem{zhang2018attention}
Y.~Zhang, J.~Du, Z.~Wang, J.~Zhang, Y.~Tu, Attention based fully convolutional
  network for speech emotion recognition, in: Proc. 2018 Asia-Pacific Signal
  and Information Processing Association Annual Summit and Conference (APSIPA
  ASC), IEEE, 2018, pp. 1771--1775.

\bibitem{zhang2016end}
S.-X. Zhang, Z.~Chen, Y.~Zhao, J.~Li, Y.~Gong, End-to-end attention based
  text-dependent speaker verification, in: Proc. 2016 IEEE Spoken Language
  Technology Workshop (SLT), IEEE, 2016, pp. 171--178.

\bibitem{chorowskiattention}
J.~K. Chorowski, D.~Bahdanau, D.~Serdyuk, K.~Cho, Y.~Bengio,
  \href{https://proceedings.neurips.cc/paper/2015/file/1068c6e4c8051cfd4e9ea8072e3189e2-Paper.pdf}{Attention-based
  models for speech recognition}, in: C.~Cortes, N.~Lawrence, D.~Lee,
  M.~Sugiyama, R.~Garnett (Eds.), Advances in Neural Information Processing
  Systems, Vol.~28, Curran Associates, Inc., 2015.
\newline\urlprefix\url{https://proceedings.neurips.cc/paper/2015/file/1068c6e4c8051cfd4e9ea8072e3189e2-Paper.pdf}

\bibitem{vaswani2017attention}
A.~Vaswani, N.~Shazeer, N.~Parmar, J.~Uszkoreit, L.~Jones, A.~N. Gomez,
  {\L}.~Kaiser, I.~Polosukhin, Attention is all you need, in: Advances in
  Neural Information Processing Systems, 2017, pp. 5998--6008.

\bibitem{sawyer2010numbers}
J.~Sawyer, By the numbers: Disfluency analysis for preschool children who
  stutter, in: Proc. International Stuttering Awareness Day Online Conference,
  2010.

\end{thebibliography}

\end{document}